\documentclass[sigconf]{acmart}
\AtBeginDocument{%
  }

\copyrightyear{2026}
\acmYear{2026}
\setcopyright{cc}
\setcctype{by}
\acmConference[CHI '26]{Proceedings of the 2026 CHI Conference on Human Factors in Computing Systems}{April 13--17, 2026}{Barcelona, Spain}
\acmBooktitle{Proceedings of the 2026 CHI Conference on Human Factors in Computing Systems (CHI '26), April 13--17, 2026, Barcelona, Spain}
\acmPrice{}
\acmDOI{10.1145/3772318.3790825}
\acmISBN{979-8-4007-2278-3/2026/04}

\usepackage{tikz}
\usepackage{tikz-timing}
\usepackage{geometry}
\usetikzlibrary{positioning}
\usepackage{multirow}
\usepackage{graphicx}

\usepackage{array}

\usepackage{balance}       
\usepackage{graphics}      
\usepackage{textcomp}
\usepackage{enumitem}
\usepackage{color}
\usepackage{epigraph}
\usepackage{subfig}
\usepackage{longtable}
\usepackage{float} 
\usepackage{microtype} 
\newcommand{\pnum}{}

\newcommand{\inlinequote}[2]{\textit{``#1''} (#2)}

\usepackage{xcolor}
\usepackage[normalem]{ulem}
\newcommand\change[1]{\textcolor{black}{#1}}

\begin{document}

\title{``It's Messy...But I Feel Balanced'': Unpacking Flexible Worker's Rhythm-Making Practices Using Asset-Based Approach}

\author{Tse Pei Ng}
\orcid{0009-0006-4339-5892}
\affiliation{%
  \department{Division of Industrial Design}
  \institution{National University of Singapore}
  \city{Singapore}
  \country{Singapore}
}
\email{tsepei@nus.edu.sg}

\author{Daniel Campos-Muñiz}
\orcid{0009-0007-1378-2768}
\affiliation{%
  \department{Division of Industrial Design}
  \institution{National University of Singapore}
  \city{Singapore}
  \country{Singapore}
}
\email{daniel_cm@u.nus.edu}

\author{Yiyang He}
\orcid{0009-0008-2540-4497}
\affiliation{%
  \department{Department of Sociology and Anthropology}
  \institution{National University of Singapore}
  \city{Singapore}
  \country{Singapore}
}
\email{heyiyang@u.nus.edu}

\author{Ker Wey Aw}
\orcid{0009-0003-7734-0521}
\affiliation{%
  \department{Division of Industrial Design}
  \institution{National University of Singapore}
  \city{Singapore}
  \country{Singapore}
}
\email{kaw@u.nus.edu}

\author{Jung-Joo Lee}
\orcid{0000-0002-7414-1936}
\affiliation{%
  \department{Division of Industrial Design}
  \institution{National University of Singapore}
  \city{Singapore}
  \country{Singapore}
}
\email{jjlee@nus.edu.sg}

\author{Janghee Cho}
\orcid{0000-0002-3193-2180}
\affiliation{%
  \department{Division of Industrial Design}
  \institution{National University of Singapore}
  \city{Singapore}
  \country{Singapore}
}
\email{jcho@nus.edu.sg}

\renewcommand{\shortauthors}{Ng et al.}

\begin{abstract}
Flexible work is increasingly pursued as a means of achieving work-life balance, particularly as growing caregiving responsibilities for children and aging family members shape workers’ lives. Yet most HCI research has examined flexibility primarily through productivity and organizational perspectives, with less attention to how it intersects with workers’ personal and family responsibilities. To address this gap, we conducted a qualitative study with 20 workers in Singapore engaging in flexible arrangements to manage paid work and care responsibilities. Using an asset-based lens, we show that flexibility is not a static benefit but a continual practice of rhythm-making. Participants maintained rhythms by drawing on temporal and spatial assets, negotiated them through relational and institutional dynamics, and sustained them through intrapersonal assets such as self-care and positive reframing. Our study reframes blurred boundaries as resources rather than disruptions and offers design implications for technologies that support flexible workers’ everyday rhythm-making practices.
\end{abstract}



\maketitle

\section{Introduction} 
Work is a socio-cultural construct that evolves in relation to economic, social, and cultural contexts~\cite{meda2016future, morse1955function}. Technological advancements have enabled new modes of work~\cite{jarrahi2021flexible, messenger2016three}, while economic needs and public health crises (e.g., the COVID-19 pandemic) have accelerated the adoption of remote and hybrid arrangements~\cite{bailey2002review, marcus2023covid}. These shifts are further shaped by demographic and societal changes, including growing caregiving demands for children and aging family members, alongside policy interventions that emphasize inclusivity and work–life balance~\cite{white2010flexible, tal2024fwa}. In response, flexible work has become a common mode of labor, offering workers autonomy over when they work, where they work, and how much they work~\cite{kossek2011flexible, tal2024fwa}. 

HCI research has examined how technologies shape work–life balance, particularly in flexible arrangements among information and knowledge workers whose practices rely on mobility, collaboration, and digital infrastructures~\cite{beckman2020dreams, erickson2016infra, sefidgar2024nonwork, vega2023platform, das2023focused}. Much of this work, however, has prioritized the work side of flexibility (e.g., optimizing productivity, coordination, and collaboration) while giving less attention to how flexible arrangements intersect with \change{domestic} roles. Although some studies have examined boundaries between home and work, these often frame \change{domestic} practices primarily in terms of their impact on paid-work~\cite{breideband2022rhythm, cho2022topophilia, williams2023boundary, sefidgar2024nonwork}. As a result, the design of interactive systems often continues to focus largely on formal work practices, with limited attention to how technologies might account for the fluid and relational dimensions of flexible workers’ everyday lives. Such orientations could risk overlooking the central role of care and domestic responsibilities, along with the invisible efforts they involve, in sustaining flexible work~\cite{cho2024feminist, steup2022attune, mckie2002shadow, de2025living}.

Historically, flexible work arrangements were concentrated among high-skilled professionals with resources to manage autonomy and work–life balance~\cite{kossek2023making, bailey2002review, cho2022topophilia}. Today, however, technological developments and government interventions have broadened access to flexibility, making it more equitable and diverse~\cite{kossek2011flexible, kossek2025reenvisioning}. For example, in Singapore, national guidelines on Flexible Work Arrangement Requests \change{were recently introduced as an initial step toward} formalizing employees’ right to request flexibility and requiring employers to give due consideration~\cite{tal2024fwa}. Such policies reframe flexibility from a privilege of the few to a shared responsibility among state, employers, and workers, balancing productivity with caregiving and well-being. As flexible work moves beyond privileged professional groups, it becomes increasingly important to recognize that its viability depends not only on organizational efficiency but also on the often-invisible efforts of workers to coordinate care and domestic life. This makes caregiving responsibilities a particularly salient dimension of flexible work and motivates our central research question: \emph{How do individuals navigate and sustain flexible work arrangements while managing ongoing caregiving responsibilities?}

To answer this question, we conducted a qualitative study with 20 participants in Singapore who engaged in flexible work arrangements while carrying out caregiving responsibilities. Drawing on an asset-based approach~\cite{jarrahi2022digital, pei2019we}, we focus on workers’ capacities to navigate multifaceted roles and blurred boundaries. \change{To move beyond surface-level accounts of work–care tensions and engage more deeply with the mundane, situated nature of these practices, we employed design probes~\cite{mattelmaki2006design} alongside semi-structured interviews. This methodological and theoretical configuration} foregrounds the strengths and strategies through which workers cultivate routines and coping mechanisms that preserve a sense of normalcy and resilience~\cite{shin2021cloud, semaan2019routine, cho2024reinforcing}. We found that flexible workers’ life priorities and care orientations shaped their approaches to rhythm-making. Rather than treating blurred boundaries solely as problems to be resolved, they navigated them as rhythms by maintaining them with temporal and spatial assets, negotiating them through relational and institutional dynamics, and sustaining them through intrapersonal assets such as self-care.

We offer three contributions to HCI research. First, we extend understandings of flexible work by presenting a probe-based situated account of how individuals negotiate flexibility alongside care and domestic responsibilities. Our participants represent a diverse set of flexible workers, moving beyond the traditional focus on office-based knowledge work. Through this lens, we highlight practices and values that remain invisible in dominant accounts of flexibility. Second, we reframe blurred boundaries not as failures or sources of spillover stress but as rhythms to be sustained, extending boundary work literature with an asset-based perspective. Third, we propose design implications for interactive systems that support rhythm-making as an ongoing practice. Rather than resolving the messiness of flexibility, we discuss how design can embrace and support it as a generative strength in everyday work and care arrangements.


\section{Related Work} 

In this section, we review three strands of literature.  We first trace the discourse on supporting work-life balance, highlighting how flexible work has emerged as a key mechanism for achieving balance and the role technologies have played in enabling it. We then examine how prior research has engaged with the intersections of paid work and life, particularly through boundary work theories~\cite{zerubavel1993fine, nippert2008home}. Finally, we discuss the asset-based approach as our analytical lens, and highlight how prior work has applied this perspective to understand resilience and everyday practices.

\subsection{Work-Life Balance and Flexible Work Arrangements}

Work–life balance (WLB) has long been a central concern in HCI, as it captures how people navigate the interplay between \change{waged labor and domestic}  spheres and how technologies mediate these dynamics, shaping productivity and well-being~\cite{teevan2022microsoft, fleck2015balancing, beckman2020dreams}. Although definitions vary, WLB generally refers to managing multiple roles, associated tensions, and satisfaction in terms of time, behavior, and fulfillment~\cite{kalliath2008work, greenhaus1985sources, poelmans2008achieving}.  \change{In this framing, ``life'' encompasses a wide range of non-waged forms of labor, including caregiving, domestic responsibilities, and other personal activities.} Early discourse on WLB emphasized managerial goals of boosting productivity through family-friendly policies~\cite{walton1973quality}, while later scholarship highlighted relational perspectives in which resources from one role could positively affect the other~\cite{voydanoff2004effects, greenhaus2006work}. More recently, the COVID-19 pandemic underscored the importance of sustaining meaningful work alongside caregiving and life beyond employment~\cite{bernhardt2023work}.  




Flexible work arrangements (FWAs) are among the most visible mechanisms for achieving WLB, encompassing flexibility in schedules, location, and workload~\cite{allen2013work,tal2024fwa}. Enabled by technological advances and supported by evolving policies, these arrangements have expanded from early forms such as telework and nomadic/mobile work to contemporary modes of remote, hybrid, and platform work, making flexibility accessible to a broader set of workers~\cite{messenger2016three, bailey2002review, erickson2016infra}. Flexible work is often associated with higher job satisfaction and productivity, particularly when workers retain control over their boundaries and operate within supportive organizational norms, which in turn fosters well-being~\cite{bailey2002review, allen2013work, kossek1999bridging}. 

At the same time, flexible work carries double-edged effects. The perception that remote work enables employees to “work from anywhere” can generate expectations of constant availability, which undermines autonomy~\cite{mols2021always, mazmanian2013autonomy}. Such expectations blur temporal boundaries between work and life, making it harder for workers to disconnect and leading to longer hours, elevated stress, and risks to overall well-being~\cite{beckel2022telework, felstead2017assessing, gajendran2007good}. Additionally, flexible workers might face other burdens, such as the emotional load of self-managing performance, social isolation, and stress from tracking tools~\cite{grant2013exploration, alvarez2023understanding}; and additional invisible work from navigating platform features that constrain their autonomy, particularly for platform freelancers~\cite{vega2023platform}. Previous literature highlight flexibility not as inherently beneficial but contingent on how workers navigate and sustain their work and life domains. This recognition has led HCI researchers to investigate how technologies might mitigate the downsides of flexibility while enabling more sustainable forms of WLB and supporting workers' overall well-being.

\subsubsection{Technologies for Flexible Workers Work-Life Balance}

One line of HCI research has focused on time management systems that help workers allocate effort and avoid overwork. These include tools that set aside protected time for focused work~\cite{saha2023time,das2023focused} and systems that encourage well-being through reminders to take breaks or switch off~\cite{das2022two, cambo2017break,epstein2016taking,howe2022microbreak}. While these tools can support well-being, they are often designed from an output-oriented perspective, emphasizing productivity and rigid temporal boundaries~\cite{guillou2020your, erickson2023optimizing}. In practice, flexible workers negotiate boundaries more fluidly, adapting to contextual demands~\cite{erickson2019flexible}. \citet{steup2022attune} describes this as \emph{attunement work}, where individuals sustain rhythms by anchoring routines, decoupling when spontaneity is needed, or blending both strategies to construct values-aligned practices.  

Beyond time management, flexible workers depend on various technical and social infrastructures to sustain their work. These tools support communication, coordination, and co-production, enabling engagement with peers remotely and asynchronously~\cite{jarrahi2021flexible, olson2000distance}. In contexts without dedicated IT support, workers must build and sustain their own infrastructures by combining digital platforms with everyday mundane technologies like splitters and peripherals~\cite{erickson2016infra, erickson2019flexible}. These digital assemblages are shaped by both workers’ personal assets and organizational provisions, including tools mandated by organizational policy that may not integrate well with workers’ existing systems~\cite{erickson2019flexible, jarrahi2022digital}. Beyond sustaining productivity, workers also pursue relational and career-development goals by leveraging technologies to make themselves visible to managers and colleagues~\cite{Mehrvarz2025visibility}. Yet these approaches remain centered on the work domain, while workers’ performance are  equally shaped by resources and demands from their family domain~\cite{greenhaus2006work, voydanoff2004effects}. This raises the question of whether flexible work infrastructures, when designed solely from the perspective of work, adequately account for how caregiving and domestic roles shape everyday practices.

Recent research highlights how flexible workers also manage caregiving responsibilities, which introduce additional coordination demands within the household–balancing schedules, managing disruptions, and aligning with the needs of care recipients and other family members~\cite{kabir2025balancing, schorch2016designing}. These demands impact workers emotionally, physically, and financially, reducing the well-being benefits expected from flexible work arrangements~\cite{duxbury2009balancing}. Moreover, blurred boundaries between home and work demand greater effort to demarcate and sustain balance~\cite{hsu2024dancing, kabir2025balancing}. With demographic shifts increasing the prevalence of care responsibilities, recent studies call for more holistic understanding of how work and caregiving spheres are intertwined. Building on this, our study examines how flexible workers navigate these intertwined responsibilities and sustain work–life balance in practice.

\subsection{Boundary Work Between Work and Life}
Designing technology for flexible workers requires understanding how individuals negotiate work and non-work roles.  Multiple lenses have been used to examine this, including border theory~\cite{clark2000work}, role transition~\cite{ashforth2000all}, and boundary theory~\cite{nippert2008home, zerubavel1993fine}. We focus here on boundary theory, which foregrounds how people actively construct, maintain, and reshape boundaries under conditions of flexibility.


Boundary theory emphasizes that people delineate temporal, spatial, and social boundaries to distinguish professional and personal domains~\cite{zerubavel1993fine, nippert2008home}. Individuals may approach boundaries with a \emph{rigid mind}, favoring clear-cut separations between domains; a \emph{fuzzy mind}, accepting blurred overlaps between spheres; or a \emph{flexible mind}, which allows them to adapt between the two depending on context~\cite{zerubavel1993fine}. In practice, these boundaries are built through everyday practices and choices, shaping the relations with the elements we interact with. Through these everyday practices, referred to as boundary work, workers construct home and work not merely as abstract or physical realms, but as experiential realities that are continually reshaped depending on circumstances~\cite{nippert2008home}. For flexible workers, this negotiation is part of everyday life, as they navigate blurred temporal and spatial boundaries, adjusting how they engage with roles~\cite{cho2022topophilia, ciolfi2020making, mckie2002shadow}.

Prior work highlights diverse forms of boundary work.  \citet{thomson2013information} describes three types of boundary work by remote workers working at home: psychological, physical, and temporal. The boundaries set by workers can range from segmented, where the separation between home and work is clear and rigid, to integrated, where the boundary is flexible allowing workers to transition more easily between them~\cite{ashforth2000all}. Although workers may establish rituals to transition between domains, these transitions can also occur unexpectedly due to interruptions, such as receiving a work email at home, blurring the boundaries between work and personal life, requiring workers to engage in micro-boundary strategies to sustain balance~\cite{cecchinato2015micro}. These strategies have become increasingly important in the context of mobile technologies that foster an “always on’’ culture~\cite{mols2021always, mazmanian2005crackberries}, paradoxically undermining workers’ autonomy over personal time~\cite{mazmanian2013autonomy}. HCI research has further shown how technologies mediate these boundaries, for instance, using separate devices for professional versus personal activities~\cite{thomson2013information, fleck2015balancing, salazar2001building}, configuring distinct applications for work and personal email~\cite{cecchinato2015micro}, or leveraging messaging app features like muting groups~\cite{mols2021always}. To mitigate these effects, researchers have developed tools such as micro-break suggestion apps that help disconnection~\cite{das2022two, cambo2017break, epstein2016taking, howe2022microbreak}. While this body of work highlights a wide variety of strategies, it often reinforces a segmentation-integration binary that oversimplifies the more fluid, and rhythmic ways to navigate boundaries.

The concept of rhythms here refers to the temporal and spatial patterns through which workers organize their work activities~\cite{nilsson2005negotiated, reddy2002pulse}. To fulfill their tasks, workers negotiate their individual rhythms with the collective rhythms of collaborative work, and the social rhythms that delineate their daily interactions~\cite{nilsson2005negotiated,sun2024mixed}. The complexity of this arrangement depends on the worker’s situatedness and organizational context. For instance, some routines are shaped by clock rhythms, defined around fixed schedules or rigid daily routines~\cite{reddy2002pulse}; or aligned with natural rhythms, adapting to bodily cycles and relational demands~\cite{leshed2014farm, steup2022attune}. While these temporal and spatial rhythms are maintained over time, they do not remain static, but rather vary and are adapted to the context. This approach highlights how spatial and temporal aspects of routines are lived together, rather than separately, which allows for a more comprehensive understanding of remote workers’ experiences ~\cite{nilsson2005negotiated}. 

In our work, we extend the study of flexible workers’ boundary work by exploring how care responsibility serves as a key driver shaping their practices and rhythms. We conceptualize boundaries not merely as constraints to be managed, but as resources that workers actively leverage in their daily lives~\cite{cho2022topophilia, bodker2016rethinking}. This perspective motivates our adoption of an asset-based approach as an analytical lens, highlighting how workers draw on their personal, technical, and organizational resources to sustain their dual roles.

\subsection{Asset-based Approach in HCI and Flexible Work}
The asset-based approach offers an alternative to deficit-oriented perspectives that emphasize the needs and problems of a community~\cite{kretzmann1996assets}. It foregrounds existing strengths and resources, recognizing them as foundations for design and grounding the idea that people can drive change themselves through their own assets~\cite{kretzmann1996assets, mathie2003clients}. In HCI, the asset-based approach has been particularly influential in studies with communities at the margins or underserved populations, highlighting the everyday strategies and resources these communities mobilize~\cite{wong2020assets, cho2019comadre, ahumada2021call, gautam2020crafting, irani2018refuge}. By anchoring interventions in existing assets, HCI researchers aim to improve the long-term sustainability of interventions, since designs do not hinge on external resources that people may have limited control over~\cite{pei2019we}. At the same time, scholars caution that focusing on assets must be accompanied by attention to context and broader systemic conditions, so as not to overlook, or even reproduce, existing power inequalities~\cite{gautam2020crafting, wong2021needs}. Thus, while the asset-based approach constitutes an alternative for designing sustainable interventions, it requires mindful application to prevent romanticization of people’s lives.

Recent studies of flexible work and caregiving have explicitly and implicitly adopted the asset-based approach. For example,~\citet{jarrahi2022digital} examined how mobile knowledge workers assemble digital and physical resources to sustain their capacity to work on the move, while~\citet{ciolfi2020making} demonstrated how home-based workers mobilize their resources to reconfigure the meaning of domestic spaces. Similarly, caregivers’ strengths have been studied, highlighting how they build networks to acquire and share knowledge~\cite{shin2021cloud}, and leverage sociomaterial resources to support themselves and others~\cite{petterson2024networks}. These studies emphasize that technology can amplify an individual’s role as a worker or caregiver, and  that any development will be embedded within  existing networks of assets. However, prior studies have examined these roles separately, leaving open questions about how workers sustain balance across these intertwined domains.

Building on previous literature on work–life balance, flexible work arrangements, and boundary work, our study adopts an asset-based perspective to examine how flexible workers navigate and fulfill both paid work and care responsibilities. Building on our central research question introduced earlier, we draw on the asset-based perspectives to further specify two lines of inquiry: 

\begin{itemize}
	\item How do individuals engaged in flexible work arrangements cultivate sustainable ways of working that align with their personal values and life priorities, particularly as they negotiate blurred boundaries between work and non-work roles? 
	\item What strategies across individual, household and relational, and organizational contexts enable or constrain these individuals in sustaining work-life balance?
\end{itemize}



\section{Methods} 
Our study sought to understand how individuals with flexible work arrangements navigate the everyday experiences of balancing paid work and caregiving responsibilities. Given the need to capture the lived realities of negotiating work and care within domestic life, we adopted a qualitative approach, using \change{design} probes~\cite{mattelmaki2006design} to surface situated routines, values, and meanings, followed by semi-structured interviews to further contextualize participants’ experiences. The study was conducted in Singapore between February and June 2025, shortly after the Tripartite Guidelines on Flexible Work Arrangement Requests came into effect in December 2024~\cite{tal2024fwa}. The study was reviewed and approved by the Institutional Review Board. Below, we detail the study context in Singapore, our participants, data collection procedures, and methods of analysis.

\subsection{Study Context} 

Singapore is experiencing rapid demographic change as it transitions into a super-aged society, increasing the need for both eldercare and childcare within households~\cite{ho2018care, wong2014paradigm}. Strong family-oriented values further make caregiving a central part of everyday life~\cite{ntuc2025familyfriendly, tan2006influence}. To mitigate these demands, Singapore has relied heavily on domestic workers, which has historically enabled dual-income households to sustain both paid work and care obligations~\cite{yeoh1999migrant}. Recently, the Tripartite Guidelines on Flexible Work Arrangements Requests were announced, mandating\change{\footnote{\textcolor{black}{While the TG-FWA mandate a process for evaluating FWA requests, they do not require employers to have FWA policies~\cite{snef2024edb}}}} that companies establish a formal process for employees to request flexible work arrangements~\cite{tal2024fwa}. This socio-cultural context positions Singapore as a distinctive case: it shares demographic pressures with other developed economies (aging, labor shortages) while resonating with other parts of the Global South where family interdependence and collective caregiving remain central~\cite{gutierrez2017tango}. Understanding flexible work in Singapore therefore offers insights into how work arrangements are negotiated within a cultural setting that prioritizes family obligations and interdependence.

\subsection{Participants and Recruitment}

To recruit individuals engaged in flexible work arrangements, we first deployed an online recruitment survey containing screening questions, along with additional background and demographic items. Eligibility criteria required participants to (1) currently engage in a flexible work arrangement, (2) be actively involved in day-to-day caregiving duties for a household member, (3) be at least 21 years of age, and (4) reside in Singapore. We drew on the Singapore government’s official language in the Tripartite Guidelines on Flexible Work Arrangement Requests to define flexible work, which includes flexi-time, flexi-place, and flexi-load arrangements, and adapted these categories for our screening questions. Participants were also required to use digital devices for most of their work, providing a common ground for examining flexible work practices in digitally mediated contexts~\cite{breideband2022teamwork}. Our focus was not on a particular occupation, but rather on understanding what has been labeled as flexible work in the Singapore context, following a practice perspective that examines how workers enact flexibility in their professional lives~\cite{erickson2019flexible}. This criterion was used to intentionally exclude common forms of work in Singapore, such as traditional home-based businesses without caregiving responsibilities and migrant platform gig work in delivery or domestic services, as these do not align with our focus on flexible work with caregiving responsibilities. Residing in Singapore was necessary for our study design, as it involved deploying physical probe kits and conducting on-site or virtual home visits during interviews. 

Participants were recruited through multiple channels, including online research participation communities (e.g., SG Research Lobang Telegram Channel\footnote{\url{https://t.me/SGResearchLobang}}) and university bulletins accessible to students, staff, and alumni. We also used snowball sampling by asking participants to refer others they knew who might be interested. These avenues were selected to reach a diverse range of participants in terms of age, occupation, and caregiving situations. We initially recruited 34 participants. However, only 22 participants completed the study\footnote{Noted that two participants (P11 and P18) are not included in our final dataset because, although both completed the probe activities and provided consent, they withdrew from the interviews due to their busy schedules.}. Twelve withdrew before completion, either declining consent after learning more about the study or withdrawing during the probe phase due to the demands of balancing caregiving and work responsibilities. Following the interviews, two participants were deemed ineligible: one (P14) was only occasionally involved in caregiving responsibilities compared with the other participants, and another (P17) was engaged in voluntary activities for a non-profit organization rather than paid flexible work. This resulted in a final sample of 20 participants (Table~\ref{tab:participants}).

\begin{table*}[htbp]
\centering
\caption{Summary of participants}
\footnotesize
\label{tab:participants}
\renewcommand\arraystretch{1.2}{
 \begin{tabular}{m{0.2cm}m{0.9cm}m{2.8cm}m{2.8cm}m{2.4cm}m{4cm}} 
\hline
\textbf{ID} & \textbf{Age and Gender}  & \textbf{Job Description} & \textbf{Work Arrangement} & \textbf{Care Responsibilities} & \textbf{Household Context} \\
\hline
P1 & 35-44, F & Part-time parenting community advocate; and freelance parenting coach & Flexi-time, Flexi-place (anywhere) & Childcare (N=2), with special needs & HDB flat; Single parent; Lives with elderly mother and aunt, 2 children, 2 live-in domestic workers (N=7) \\
\hline
P2 & 35-44, M & Full-time procurement lead at an MNC & Flexi-time, Flexi-place (Hybrid) & Childcare (N=2) & Condo; Lives with spouse, 2 children and live-in domestic worker (N=5) \\
\hline
P3 & 35-44, F & Full-time sustainability consultant at an MNC & Flexi-time, Flexi-place (Hybrid) & Childcare (N=2) & Condo; Lives with spouse, 2 children and live-in domestic worker (N=5) \\
\hline
P4 & 35-44, F & Part-time operations executive; and freelance Pilates teacher (home-based) & Flexi-time, Flexi-place (Anywhere) & Childcare (N=3) & HDB flat; Lives with spouse, 3 children and live-in domestic worker (N=6) \\
\hline
P5 & 45-54, F & Contracted adjunct teacher at a public school & Flexi-time, Flexi-place (Hybrid) & Eldercare (N=2); adult daughter with medical condition (N=1) & HDB flat; Lives with spouse, 2 elderly parents, 3 children and live-in domestic worker (N=8) \\
\hline
P6 & 25-34, F  & Co-founder of a social enterprise & Flexi-time, Flexi-place (Anywhere) & Eldercare (N=1); Supporting care for grandmother (N=1) & Landed Housing; Lives with parents, sibling and live-in domestic worker (N=5) \\
\hline
P7 & 35-44, F  & Full-time finance staff & Flexi-time, Flexi-place (Anywhere) & Eldercare (N=2); Childcare (N=2) & HDB flat; Single parent; Lives with mother, father-in-law, and 2 children (N=5) \\
\hline
P8 & 35-44, F & Full-time healthcare staff at a hospital & Flexi-time (staggered hours), Office-based (some hybrid) & Childcare (N=1) & Condo; Lives with spouse and 1 child (N=3) \\
\hline
P9 & 35-44, F & Self-employed realtor & Flexi-time, Flexi-place (anywhere) & Childcare (N=1), with special needs; Supporting care for nieces and nephews (N=3) & HDB flat; Lives with child, child’s father and live-in domestic worker (N=4) \\
\hline
P10 & 35-44, F  & Various part-time, ad-Hoc work (e.g. bubble tea shop assistant, mystery shopper) & Flexi-time, Flexi-place (Anywhere) & Childcare (N=3) & HDB flat; Lives with spouse and 3 children (N=5) \\
\hline
P12 & 45-54, F & Permanent part-time marketing communications executive & Flexi-time, Hybrid & Childcare (N=1); with special needs & HDB flat; Lives with spouse, child and live-in domestic worker (N=4) \\
\hline
P13 & 45-54, F  & Full-time Independent surveyor & Flexi-time, Flexi-place (anywhere) & Childcare (N=1); with special needs & HDB flat; Single parent living with child (N=2) \\
\hline
P15 & 55-64, M & Contract adjunct teacher at a public school & Flexi-time (reduced hours); Office-based & Eldercare (N=1) & HDB flat; lives with elderly mom and live-in domestic worker (N=3) \\
\hline
P16 & 35-44, F & Full-time legal counsel at an MNC & Flexi-time, Hybrid & Childcare (N=2) & Landed Housing; Single parent, living with elderly parents and 2 children (N=5) \\
\hline
P19 & 55-64, F & Permanent part-time senior care specialist & Flexi-time, Office-based & Eldercare (N=1) & HDB; Lives with elderly mother (N=2) \\
\hline
P20 & 35-44, F  & Full-time graphic designer at magazine company & Flexi-time, Flexi-place (anywhere) & Childcare (N=2) & Condo; Lives with spouse and 2 children (N=4) \\
\hline
P21 & 35-44, M & Freelance (project based) graphic designer & Flexi-time, Flexi-place (anywhere) & Childcare (N=1) & HDB; Lives with spouse, child and elderly parents (N=5) \\
\hline
P22 & 35-44, F & Freelance editor for a magazine company & Flexi-time, Flexi-place (anywhere) & Childcare (N=2); Eldercare (N=2) & HDB; Lives with spouse, 2 children, elderly parents, and 2 live-in domestic workers (N=8) \\
\hline
\change{P23} & 21-24, F  & Full-time researcher at a university & Flexi-time, Hybrid & Sibling with mental health condition (N=1) & HDB; Lives with sibling (N=2) \\
\hline
\change{P24} & 21-24, F & Full-time admin staff & Flexi-time, Hybrid & Eldercare (N=1) & HDB; Lives with elderly mother (N=2) \\
\hline
\end{tabular}}
\\[1ex] 
\parbox{\textwidth}{\footnotesize
Note: F = Female, M = Male; MNC = Multi-National Corporation; HDB = public housing flats developed by the Housing and Development Board in
Singapore. They provide affordable homes for citizens and are not limited to low-income groups; Special needs = Here, we use the term “special needs” as our participants did, to refer to cognitive and physical diversity. P11, P14, P17, and P18 are not included due to eligibility criteria, regardless of whether they completed the study or did not finish all components. See Section 3.2 for details.}
\end{table*}

\subsection{Data Collection \& Procedure} 

\begin{figure*}[ht]
 \centering
  \subfloat[]{\includegraphics[width = 0.35\linewidth]{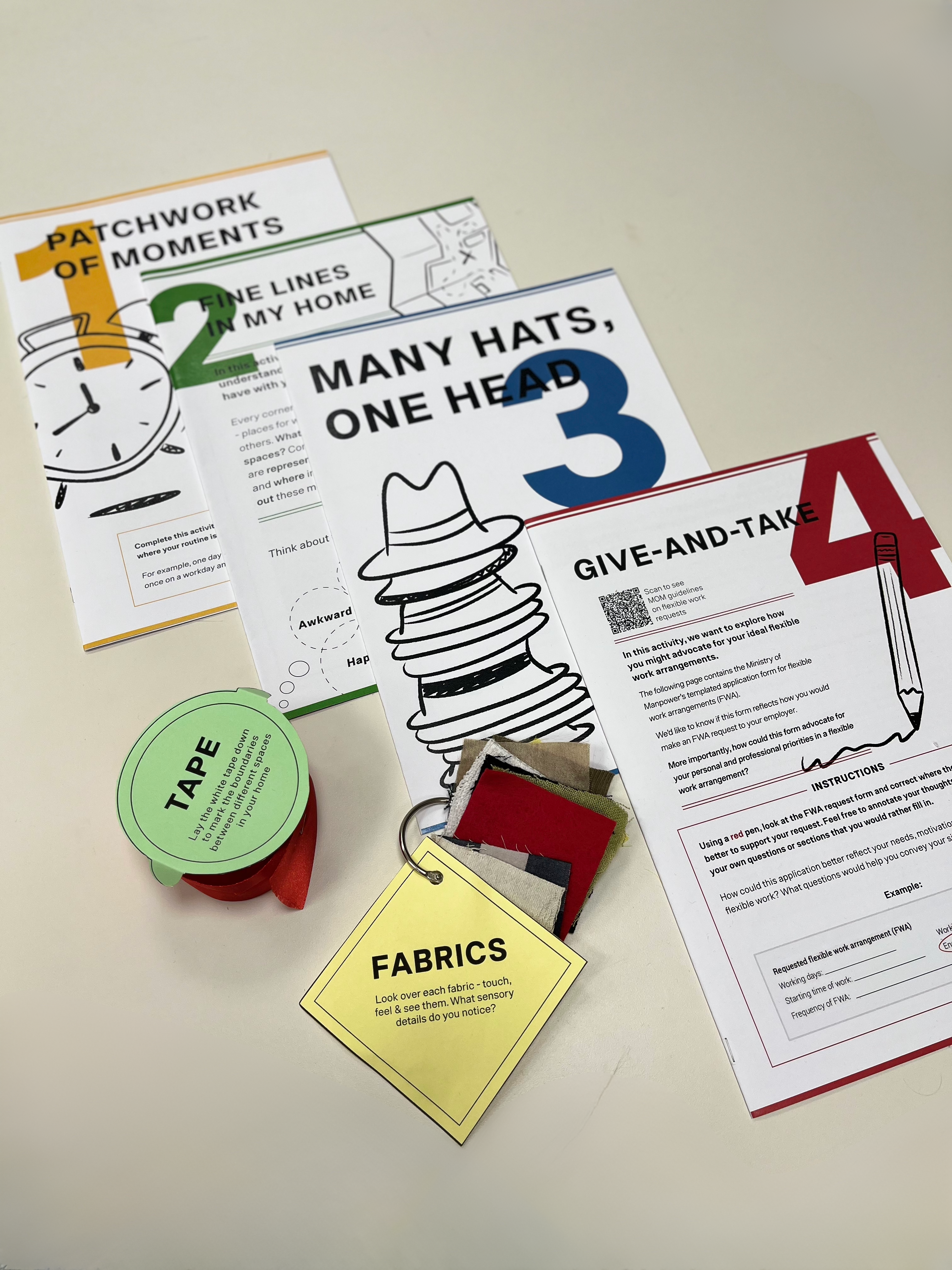}} \enskip
  \subfloat[]{\includegraphics[width = 0.62\linewidth]{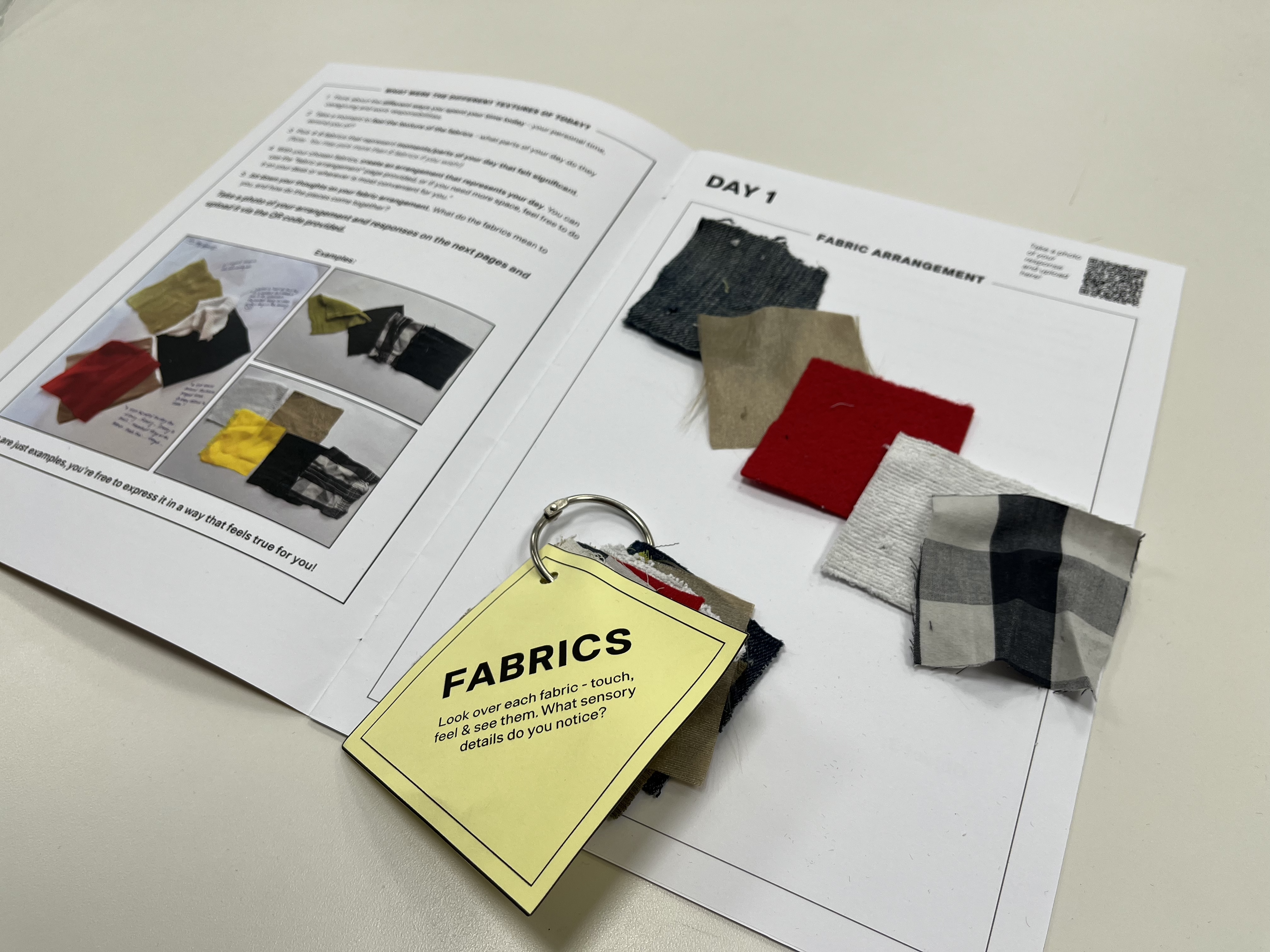}}
\caption{(a) \change{Design} probes workbook and supplementary materials; (b) Activity 1 of \change{design} probes workbook, where participants were invited to express their routines by arranging fabric pieces on the page.}
\Description{Two photos labelled (a) and (b) showing workbooks and supplementary materials for \change{design} probes. Image (a) shows a set of four workbooks with a roll of tape and set of fabrics; Image (b) shows an example of activity one, depicting an opened workbook with five fabric squares of various colours and textures arranged diagonally across the page.}
\label{fig:CulturalProbes}
\end{figure*}

After obtaining consent, we deployed a design probe kit, \change{designed with inspiration from} cultural and design probe methods~\cite{gaver1999probes, mattelmaki2006design}, consisting of a workbook containing short introductory guidelines and reflection prompts, and supplementary materials (e.g., fabric and tapes) (Figure~\ref{fig:CulturalProbes}). \change{Participants were asked to respond to four activities: (i)\emph{`Patchwork of Moments'}, where participants used fabric swatches to represent their daily routines and presented objects significant to those routines across three different days; (ii)\emph{`Fine Lines in My Home'}, where participants used tape to mark the boundaries between spaces according to the meanings they attributed to each space; (iii)\emph{`Many Hats, One Head'}, where participants reflected on the priorities associated with their roles and values, considering both their current situation and their ideal; (iv)\emph{`Give and Take'}, where participants annotated the Ministry of Manpower’s Flexible Work Arrangement request form to reflect the nature of their flexible work and to articulate the support they desired from both their employers and their care recipients. These activities were crafted to prompt reflection and open-ended responses. For example, the fabric swatches in activity one prompts participants to reflect with the textures and colours of the fabrics to express the emotional textures of their daily rhythms. Altogether, the activities intended  to capture participants' day-to-day emotional, felt experiences, the meanings they attach to daily objects and spaces, and their values and priorities.} We conducted a pilot test with individuals engaged in flexible work and caregiving responsibilities to refine the prompts and materials. Our aim was to use the probe not solely as a data collection tool, but also \change{as a sensitizing scaffold for the subsequent interviews, helping} build empathetic engagement with participants and elicit more provocative and inspirational insights for design~\cite{boehner2007intepret, wallace2013making}. The probe kits were delivered and retrieved either by mail or in person. Participants were encouraged to complete the activities within a two-week timeframe, although some required up to four weeks due to personal circumstances. 

Upon completion of the probe activities, we conducted semi-structured interviews to gain a deeper understanding of participants’ flexible work arrangements and how they managed the intersection of work and caregiving responsibilities. The completed probes \change{helped anchor the conversation and supported empathetic engagement.} Each interview lasted 60–90 minutes and was typically conducted via Zoom\footnote{Participants could choose between an in-person interview at their home or a Zoom session; however, only one participant opted for the in-home format}, during which participants could provide a brief home tour. Two research team members facilitated the interviews.

\subsection{Data Analysis}
All probe materials were digitized, including participants’ photographs and workbook responses. Interview recordings were transcribed for analysis, and all data were anonymized to remove personally identifying information. We adopted a constructivist grounded theory approach~\cite{charmaz2014constructing}, engaging in an iterative, abductive coding process. The first and last authors conducted open coding in ATLAS.ti, employing a variety of techniques (e.g., in vivo, process, descriptive, and emotion coding~\cite{saldana2015coding}) to capture emergent themes. Throughout the process, we held regular discussions to align our coding, reach consensus on code definitions, and share analytic memos that documented each researcher’s interpretations. Our research team also occasionally held full-team meetings to share interim findings, discuss emerging patterns, and refine our analytic focus. 

\change{While participants’ probe responses were not directly analyzed as an explicit stage in the analysis, they sensitized us to participant’s situated contexts, revealing the deeper values driving their actions within situations of constraint. This orientation informed multiple rounds of open coding, inspiring us to highlight participants’ creative strategies that draw on personal, relational, and material resources to navigate tensions and (re)shape rhythms.} We adopted an asset-based approach~\cite{shin2021cloud, wong2020assets} as the theoretical lens in our focused coding to examine how participants sustained rhythms of work and care. Building on this, we conceptualize rhythm-making as everyday infrastructuring:  the ongoing work of constructing, sustaining, and adapting one's own assets to manage the entanglement of paid work and caregiving. To deepen our analysis, we applied situational analysis~\cite{clarke2003situational} to situate these strategies within broader configurations (e.g., human and nonhuman actors, policies, discourses, or cultural norms), allowing us to examine both participants’ actions and the complex contexts in which they were embedded. 

\section{Findings} \label{findings}
In this section, we outline our three key findings on how participants draw on and configure a range of assets to enact and sustain flexibility. These assets constitute everyday arrangements, spanning temporal and spatial resources, relational and institutional supports, and intrapersonal capacities. Our analysis highlights how participants leverage these assets to: (1) maintain rhythms through boundary navigation, (2) negotiate rhythms through relational and institutional dynamics, and (3) sustaining rhythms through intrapersonal strategies such as self-care.

\begin{table*}[htbp]
\centering

\footnotesize
\label{tab:findings}
\renewcommand\arraystretch{1.2}{
\begin{tabular}{m{5cm}m{2.5cm}m{7cm}}
\hline
\textbf{Layer of Rhythm-Making} & \textbf{Asset Type} & \textbf{Strategy Examples}\\
\hline

Work and Care Maintenance of Rhythms (Section 4.1)
  & Temporal (Short-term) &
  \begin{minipage}[c]{\linewidth}
  \vspace{3pt}
  \begin{itemize}[leftmargin=*]
    \item ‘Soft rebalancing’ to make up for caregiving time 
    \item Multi-tasking adaptively to attend work while caring
    \item Leveraging pockets of time in-between care for work 
  \end{itemize}
  \vspace{2pt}
  \end{minipage}\\
\cline{2-3}

  & Temporal (Long-Term) &
  \begin{minipage}[c]{\linewidth}
  \vspace{3pt}
  \begin{itemize}[leftmargin=*]
    \item Deliberate long-term career tradeoffs to preserve flexibility 
  \end{itemize}
  \vspace{2pt}
  \end{minipage}\\
\cline{2-3}

  & Spatial &
  \begin{minipage}[c]{\linewidth}
  \vspace{3pt}
  \begin{itemize}[leftmargin=*]
    \item Co-locating work and care by curating overlapping environments
  \end{itemize}
  \vspace{2pt}
  \end{minipage}\\
\hline

Household-level Negotiation of Rhythms (Section 4.2.1 \& 4.2.2)
  & Relational &
  \begin{minipage}[c]{\linewidth}
  \vspace{3pt}
  \begin{itemize}[leftmargin=*]
    \item Proactive schedule coordination to distribute care responsibilities
    \item Real-time tag teaming with spouse or domestic worker
    \item Extended network buffering when household dynamics are strained
  \end{itemize}
  \vspace{2pt}
  \end{minipage}\\
\cline{2-3}
\hline

Organizational Negotiation of Rhythms (Section 4.2.3 \& 4.2.4)
  & Institutional \& Relational (Policy / Culture) &
  \begin{minipage}[c]{\linewidth}
  \vspace{3pt}
  \begin{itemize}[leftmargin=*]
    \item Activating supportive workplace policies to freely configure schedules
    \item Credibility building by requesting flexibility only when essential
    \item Mutual support to cover one another’s work
    \item Proactive impression management to maintain goodwill with colleagues
  \end{itemize}
  \vspace{2pt}
  \end{minipage}\\
\hline

Self-based Practices for Sustaining Rhythms (Section 4.3)
  & Intrapersonal &
  \begin{minipage}[c]{\linewidth}
  \begin{itemize}[leftmargin=*]
  \vspace{3pt}
    \item Micro-boundary practices to reset during work-care transitions
    \item Low effort “mindless” rest using screen-based media
    \item Cultivating helpful inner orientations amidst ongoing stress
  \end{itemize}
  \vspace{2pt}
  \end{minipage}\\

\hline
\end{tabular}}
\\[1ex]
\caption{Summary of the layers of rhythm-making and associated assets.}

\end{table*}

\subsection{Maintaining Rhythm through Temporal and Spatial Assets} \label{finding-rhythm-1}
\begin{figure*}[ht]
 \centering
  \subfloat[]{\includegraphics[width = 0.3\linewidth]{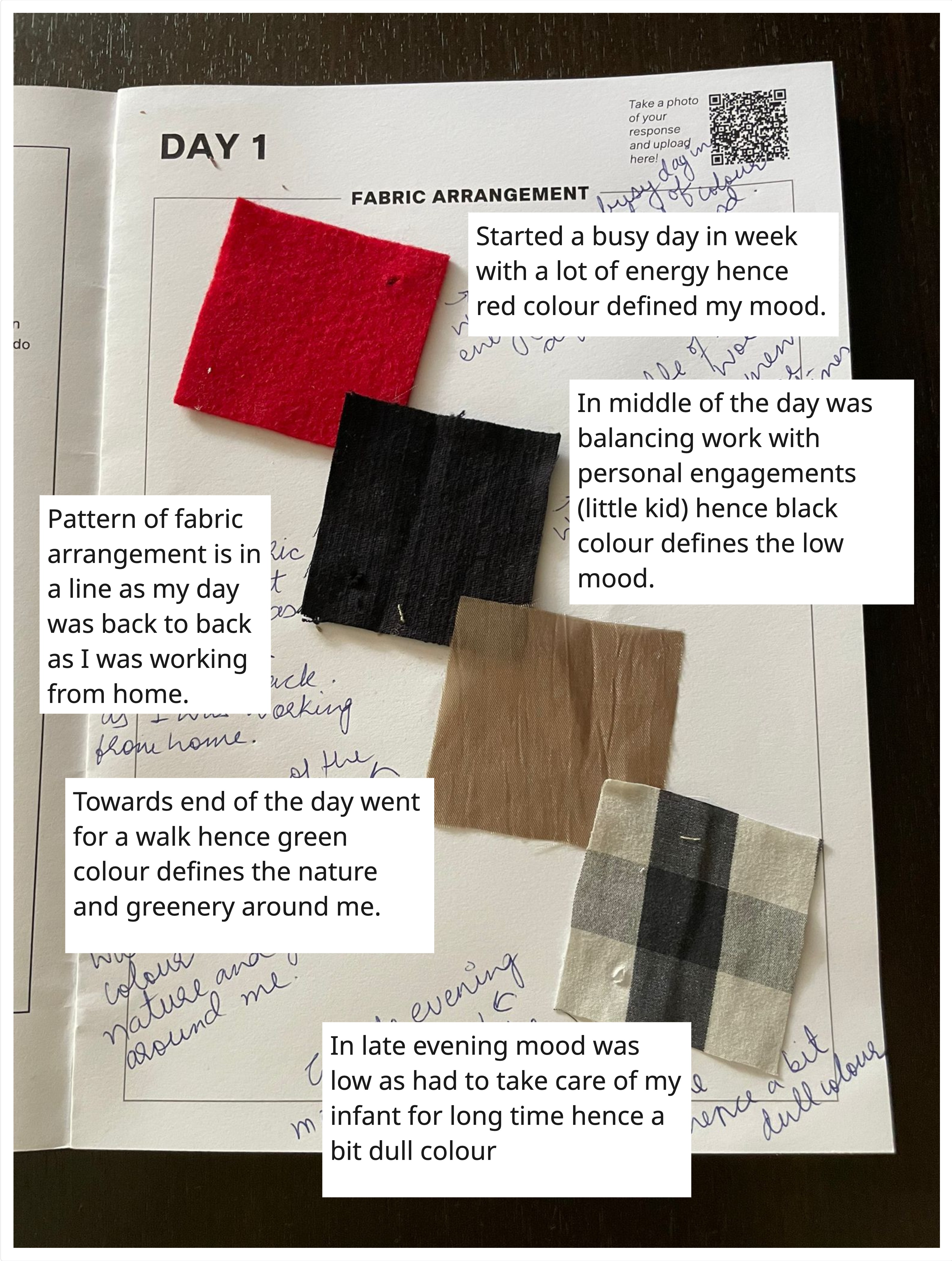}} \enskip
  \subfloat[]{\includegraphics[width = 0.3\linewidth]{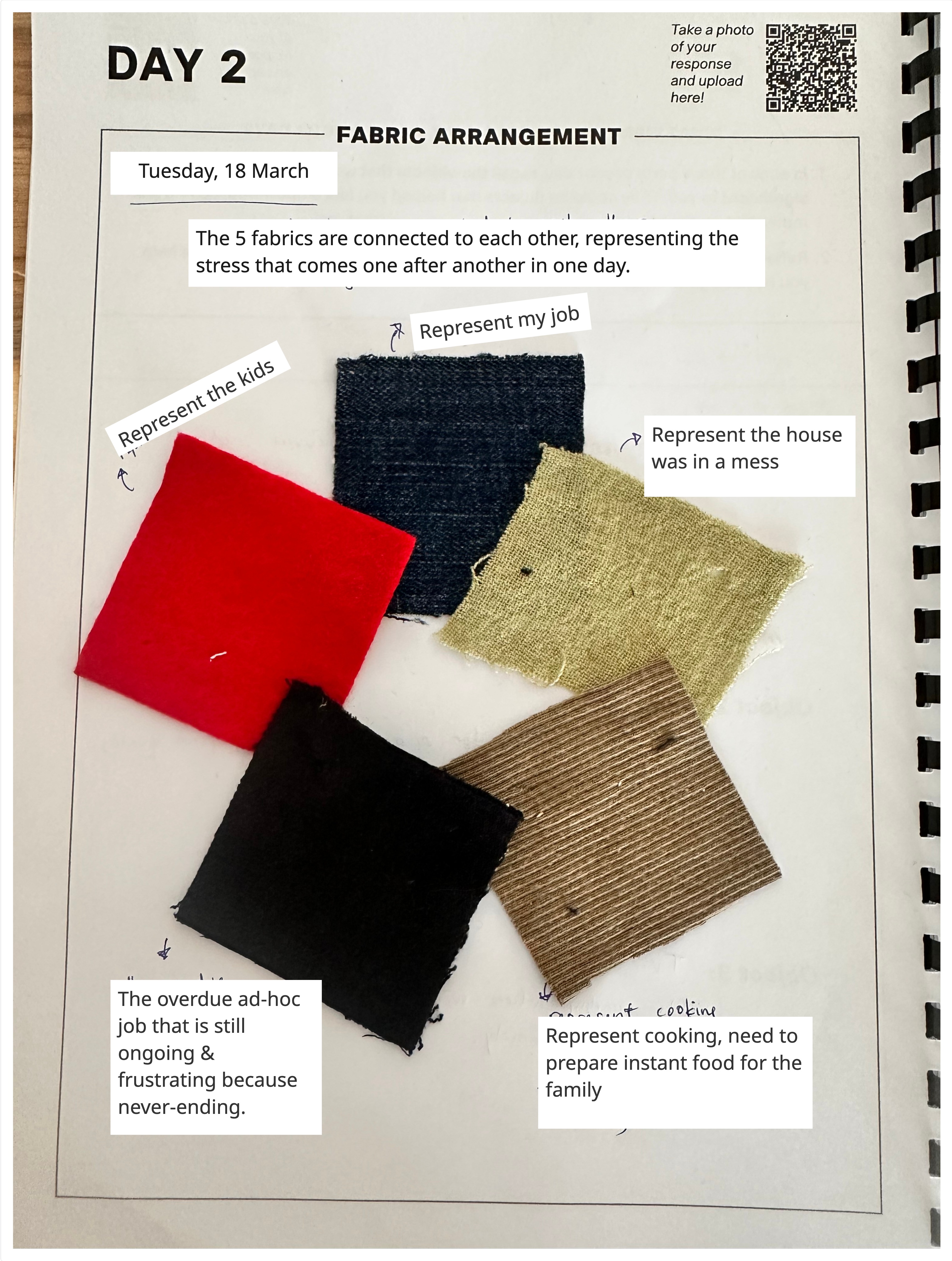}} 
\caption{Participants' probe responses \change{visualize the felt intensity of} blurred temporal boundaries between work and care: (a) P2 arranged his fabric overlapping in a line, representing his \inlinequote{back-to-back}{P2} day balancing care and work from home; (b) P20 placed 5 fabric pieces in a connecting circle to represent the \inlinequote{stress that comes one after another in one day}{P20}}. 
\Description{Two photos labelled (a) and (b), depicting participants' responses to activity one of design probes. Image (a) shows four fabric squares of red, black, brown and checkered pattern arranged diagonally across the page from top left to bottom right; image (b) shows five fabric squares of red, black, brown, ochre and dark blue, arranged in a circle.}
\label{fig:fabricArrangements}
\end{figure*}

Contrary to prior work framing blurred boundaries simply as sources of stress, our findings show that participants leveraged these blurred boundaries (visualized as overlapping layers in Figure~\ref{fig:fabricArrangements}) as critical assets. \change{While this required articulation work, participants did not view it merely as a constraint to be endured. Instead, they actively configure temporal and spatial assets to align their daily rhythms with their core caregiving values. By transforming potential conflicts into strategic permeability, they were able to sustain their dual identities as caregivers and flexible workers.} 


\subsubsection{Fine-Grained Rebalancing in Daily Routines} \label{finding-finegrained}
Participants’ daily work-care rhythms varied with their work arrangements, shaping how blurred or segmented their boundaries could be. Those in hybrid roles requiring fixed hours or on-site presence described somewhat more segmented rhythms. However, they used whatever flexibility they possessed to attend to unexpected caregiving demands, such as sudden illnesses or urgent appointments. For instance, P16, a hybrid worker at a Multinational Corporation (MNC), \change{described this as a \emph{``soft rebalancing''} shifting meetings to accommodate a sick child, then returning to work when care demands eased.} \change{Even when this resulted in near-constant availability, participants framed permeability as a value-driven trade-off. P13 acknowledged the lack of boundaries, but she utilized this total flexibility to enable her to parent more actively:} \inlinequote{You can't draw the line anywhere. Basically, you just have to be available when you're needed}{P13}. 



\change{Participants} with full flexibility (i.e., work anytime, anywhere) experienced far more blurred boundaries in both time and space. Rather than segmenting, they \change{leverage their spatial and temporal flexibility to configure interleaved rhythms, enabling them to remain responsive to emerging care needs.} 
P21, a freelancer working from home with a steady stream of projects from a long-term client, described \change{adaptively multi-tasking} by scaling his work intensity to his infant’s rhythms \change{and leveraging pockets of time between caring}: \inlinequote{When she's napping... You need to keep one eye on her…So…if I can, I'll do some admin stuff, admin stuff that don't need to think…}{P21}. \change{He further maximized pockets of time while his child was in school:} \inlinequote{…cannot let this period to go to waste…do as much as possible…super multi-tasking…before the child come back.}{P21}. Similarly, P20 tried to \inlinequote{work as much as I can during that day}{P20} when her children are at school, to cope with her overlapping work and care tasks (see more detail in Appendix\footnote{The Appendix presents extended case accounts that focus on certain participants and highlight the depth and nuance of their everyday negotiations of work and care.}~\ref{case}).

\begin{figure*}[ht]
 \centering
  \subfloat[]{\includegraphics[width = 0.3\linewidth]{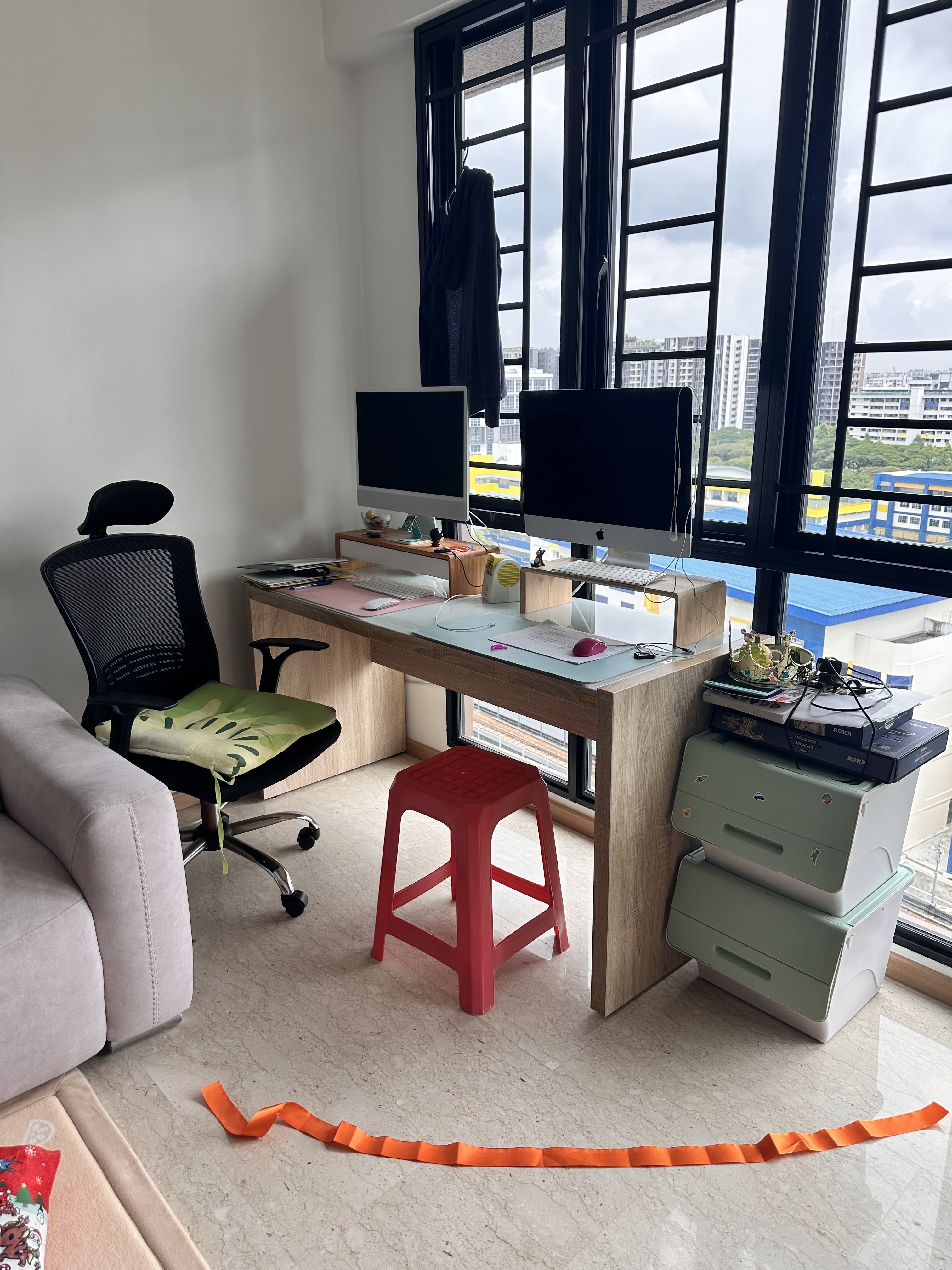}} \enskip
  \subfloat[]{\includegraphics[width = 0.3\linewidth]{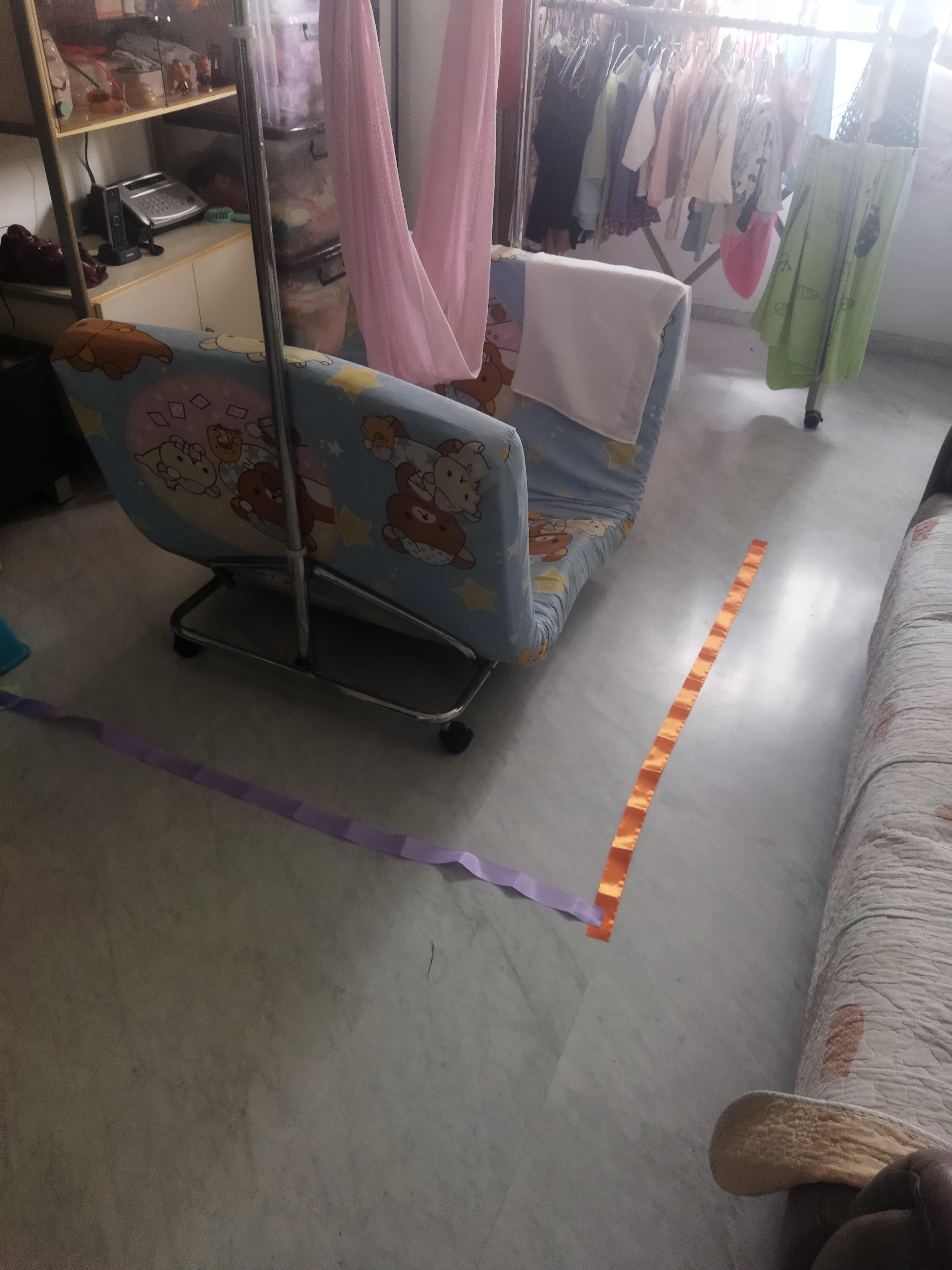}} \enskip
\subfloat[]{\includegraphics[width = 0.3\linewidth]{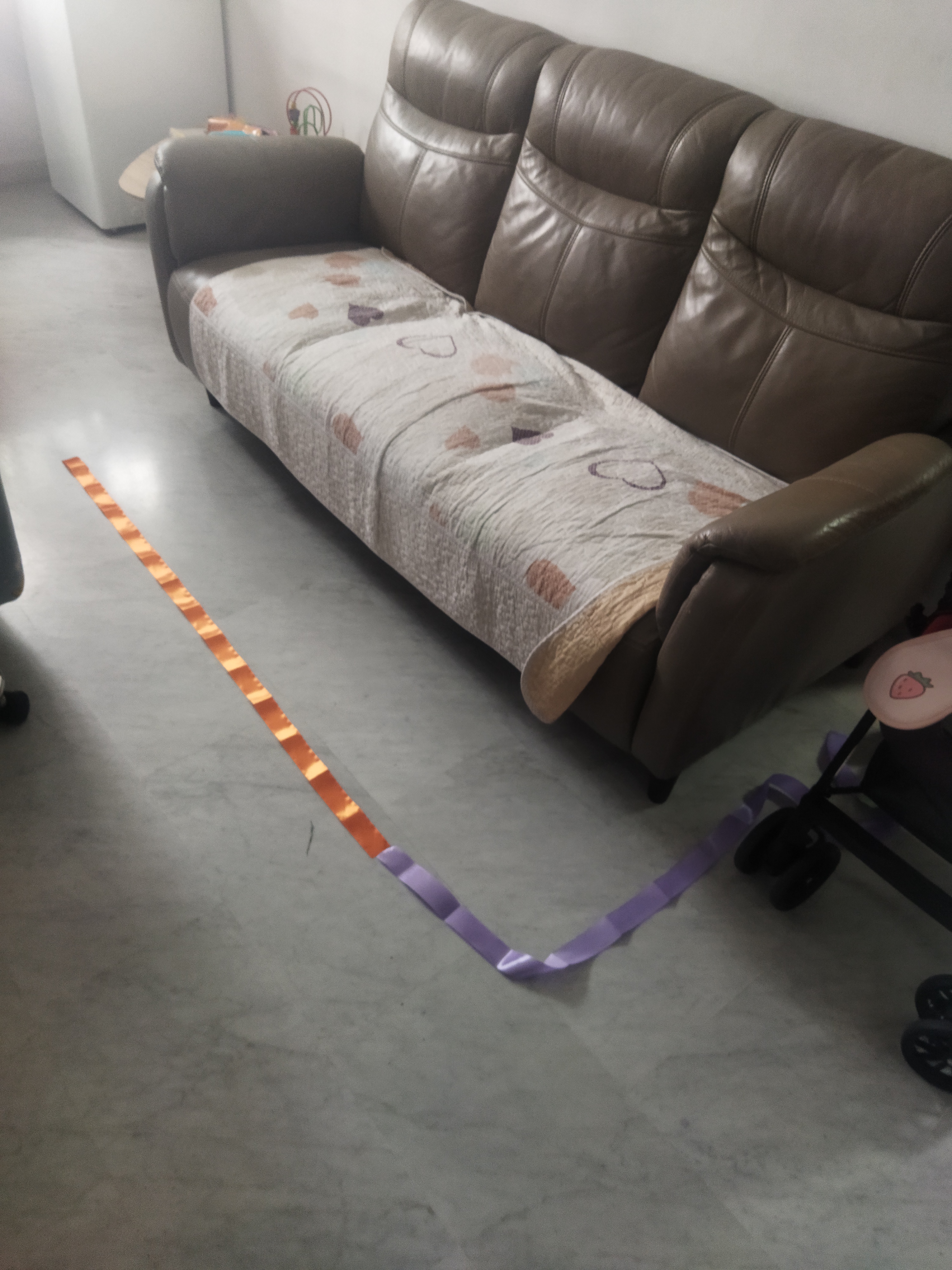}}
\caption{(a) P20’s shared work desk in her living room, \change{intentionally arranged to support simultaneous supervision and work}; (b) P21’s baby’s cradle (pink cloth) \change{positioned besided the living room, enabling continuous monitoring during work}; (c) P21’s couch next to his baby’s cradle, where he works while rocking her to sleep. \change{All photos were taken by participants as part of Design Probe Activity 2.}}
\Description{Three photos labelled (a), (b), and (c), depicting participants' home spaces. Image (a) shows a work desk in a living room, with two monitors and two chairs side by side. Image (b) shows a pink hanging cloth cradle with a mattress underneath, and image (c) shows a three seater leather couch next to it.} 
\label{fig:Spatial configurations}
\end{figure*}

\change{To support this temporal interleaving, participants also curated overlapping work and care environments (Figure \ref{fig:Spatial configurations}). For example,} P20 established a shared desk in the living room to supervise her children while working (see more detail in Appendix~\ref{case}). Likewise, P21 worked on a laptop from the couch while rocking his child’s cradle in the living room, sustaining rhythms of both care and work through the arrangement of physical settings. In another case, P4 extended spatial curation into the digital realm, scaling her work across devices—using a phone for quick replies while attending infant classes and a laptop while waiting for her older child at tuition.

\change{For these participants, the boundary permeability was not a failure of planning but a \emph{pragmatic} mode of rhythm-making, sustained through articulation work such as rescheduling meetings and toggling between communication channels to hold multiple domains together. Rather than treating spillover as a breakdown of boundary management, participants framed it as a necessary condition for sustaining livable rhythms. Blurred boundaries thus became a resource for adaptability under uncertainty, allowing them to remain accountable in professional roles while keeping caregiving at the center of their priorities. As P4 explained:}

\begin{quote}
    \textit{``I feel quite happy to be able to do that (answering work texts while caring) because like the alternative... I wouldn't even have the luxury of being there for the play date…I don't view it as like encroaching on my time that I can spend with my children, but rather it's enabling me to even spend that little bit of time with them.''} (P4)
\end{quote}


\subsubsection{Coarse-grained Prioritization and Long-term Trade-offs} \label{finding-coarse}
\change{Beyond daily adjustments, participants also negotiated rhythms at broader temporal scales.} Many participants often described their routines as messy \change{not because of poor planning, but because they oriented around caregiving priorities rather than detailed scheduling.} In this coarse-grained mode of rhythm-making, participants aligned work around what caregiving made possible, preserving flexibility for when critical needs arose. For example, P20 explained that she avoided scheduling apps \change{for time and task management}, instead responding to whatever task was most pressing: \inlinequote{I don't do any like scheduling or what…I just look at the email what needs to be done and then I'll just do… If (work) really have urgency, then I'll just focus on the work and whatever can wait.}{P20}. 

\change{This strategic prioritization extended to longer-term trade-offs to preserve temporal autonomy.} Some deferred career progression, accepted lower pay, or chose freelance arrangements despite financial risks. P1 delayed returning to work for two years to support a child with special needs, while P21 postponed full-time employment until his child adjusted to daycare: : \inlinequote{I think, uh, the child will come first. Wait for the child to be suitable…to the (daycare) place. Then, uh, then I will think about (going back to full-time)}{P21}. Likewise, P20 \change{chose to prioritize her children over career advancement despite feeling stagnant}: \inlinequote{I don't feel that I'm growing... I think it's also hard because I need the flexibility that I'm having now. So if I have to move on to upgrade myself, study or change job, I think I wouldn't be able to do what I'm doing now. Because my priority is still family and my kids... when they grow up, then I can improve on my career}{P20}.


These longer-term trade-offs were particularly common among freelancers and contract workers, who often felt more secure than in full-time employment because they faced fewer institutionalized expectations of ``work devotion’’~\cite{williams2016beyond} and could more openly prioritize caregiving. For these participants, temporal autonomy itself became a critical asset. What appeared as sacrifice in conventional career logics was reframed as an enabler of livable balance, even when daily routines seemed disorderly. As P20 reflected \inlinequote {Although (the routine is) very messy… \change{but I get to spend more time with my kids…I think I can do everything like how I’m doing now...} I feel quite balanced as well.}{P20}.

\subsection{Negotiating Rhythms through Relational and Institutional Dynamics}  \label{finding-rhythm-2}
\label{subsec:relational-dynamics}
While some rhythms could be sustained individually, most required drawing relational and institutional assets through negotiation with others. \change{Flexibility is collectively sustained; participants} coordinated with family members, extended kin, colleagues, and employers, to share or redistribute responsibilities when rhythms cannot be sustained alone, though these supports were often uneven. In this section, we examine how relational and institutional assets supported—or, when absent, strained—participants’ efforts to sustain work–care rhythms. 

\subsubsection{Negotiating Relational Dynamics within the Household} \label{finding-relational-1}
For some participants, care was jointly managed with co-caregivers in the household---most often spouses, and in some cases, live-in domestic workers. Together, they manage work and care in tightly coupled ways---syncing, reviewing and negotiating work schedules through digital tools such as calendars and messenger groups. For example, P3 syncs calendars with her husband so they could anticipate each other’s commitments.  P2 reviews work schedules weekly with his wife, to determine who will work from home and handle caregiving (see Figure~\ref{fig:fabricArrangements} (a)). \change{Others such as P10 utilized dedicated chat groups with her spouse to solicit support, track important information, dates and to-dos.}
\begin{quote}
	\textit{“...in our telegram chat window, we have one channel that is for schedule. We will slot our work or whatever event related stuff inside… So I'll know when he's busy and I have to take care of the kids and he'll know when I'm busy and he has to take care of the kids...''} (P10)
\end{quote}

Beyond planning, real-time tag-teaming helped manage contingencies and provided emotional relief, sustaining the rhythms of care and work. For P2, infant care often disrupted planned work and care tasks, requiring constant attunement and occasional reliance on his domestic worker during mandatory online meetings (see more details in Appendix~\ref{case}). Similarly, P10 and her husband both balance part-time flexible work with care for three young children, describe taking turns and adjusting as contingencies arose: 
\begin{quote}
\textit{“...in the evening…after we put them to sleep together, he goes out to work again until like 2 a.m…So (during the day) when I have mystery shopping assignment or other flexible appointments…he would look after the baby so that I can go and work.”} (P10)
\end{quote}
She also described how emotional strains triggered similar coordination. When she felt drained by late-night disruptions while caring for the children, she would text her husband for emotional support. Similarly, P8 explains, \inlinequote{sometimes if the kid has been with you the whole day, sometimes you just need to distance yourself away for some time. So that's when the other spouse comes in useful.}{P8}.


\change{In these cases, participants} treat relational coordination with their co-caregivers as an everyday asset, \change{leveraging ongoing negotiations (e.g., proactive scheduling and real-time tag-teaming) to collectively sustain work and care while gaining emotional support and respite.}

\subsubsection{When Household Support Fails: Coping Strategies Beyond the Home} \label{finding-relational-2}

When \change{relational assets within the household were} insufficient or constrained, participants turned to extended kin, community networks, and personal coping strategies. P24, an only child caring for her mother, described tensions in their relationship: \inlinequote{sometimes when she says hurtful things, right, you ask her to stop, she can't stop}{P24}. Without household members to share the emotional load, she relied on extended networks and hobbies for self-care. 

Others, \change{like P7, relied on community networks. With inconsistent support from her household members} \change{she drew from her temple community for emotional support, finding camaraderie and reframing} her struggles through shared stories: \inlinequote{sometimes we talk about our challenges… I feel like, oh, I'm not alone, then I feel my burden, it's small}{P7}. \change{When household support was limited, such external relationships became essential assets for sustaining their rhythms of work and care.}

\begin{figure*}[ht]
 \centering
  \subfloat[]{\includegraphics[angle=270, width = 0.3\linewidth]{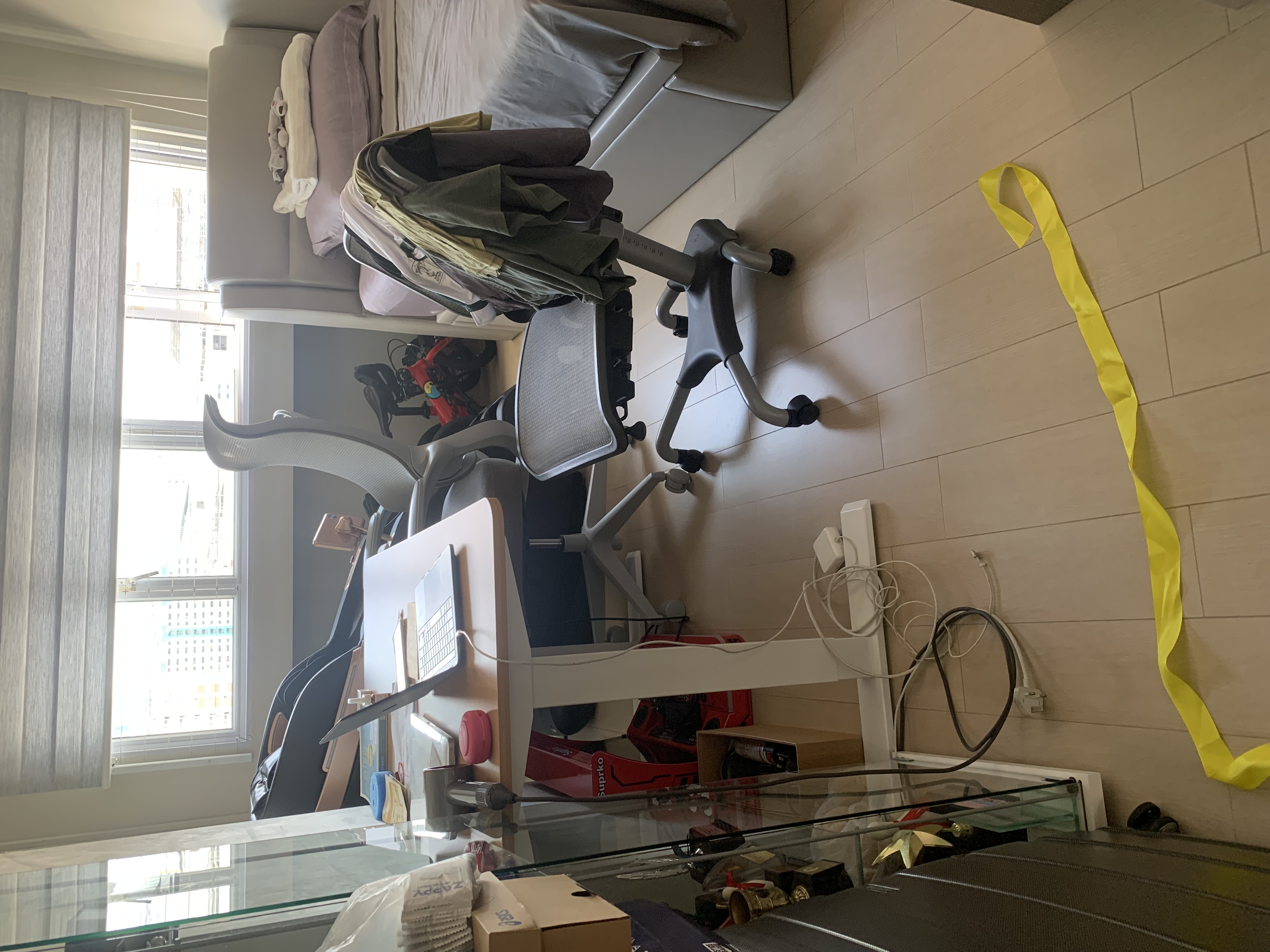}} \enskip
  \subfloat[]{\includegraphics[angle=270, width = 0.3\linewidth]{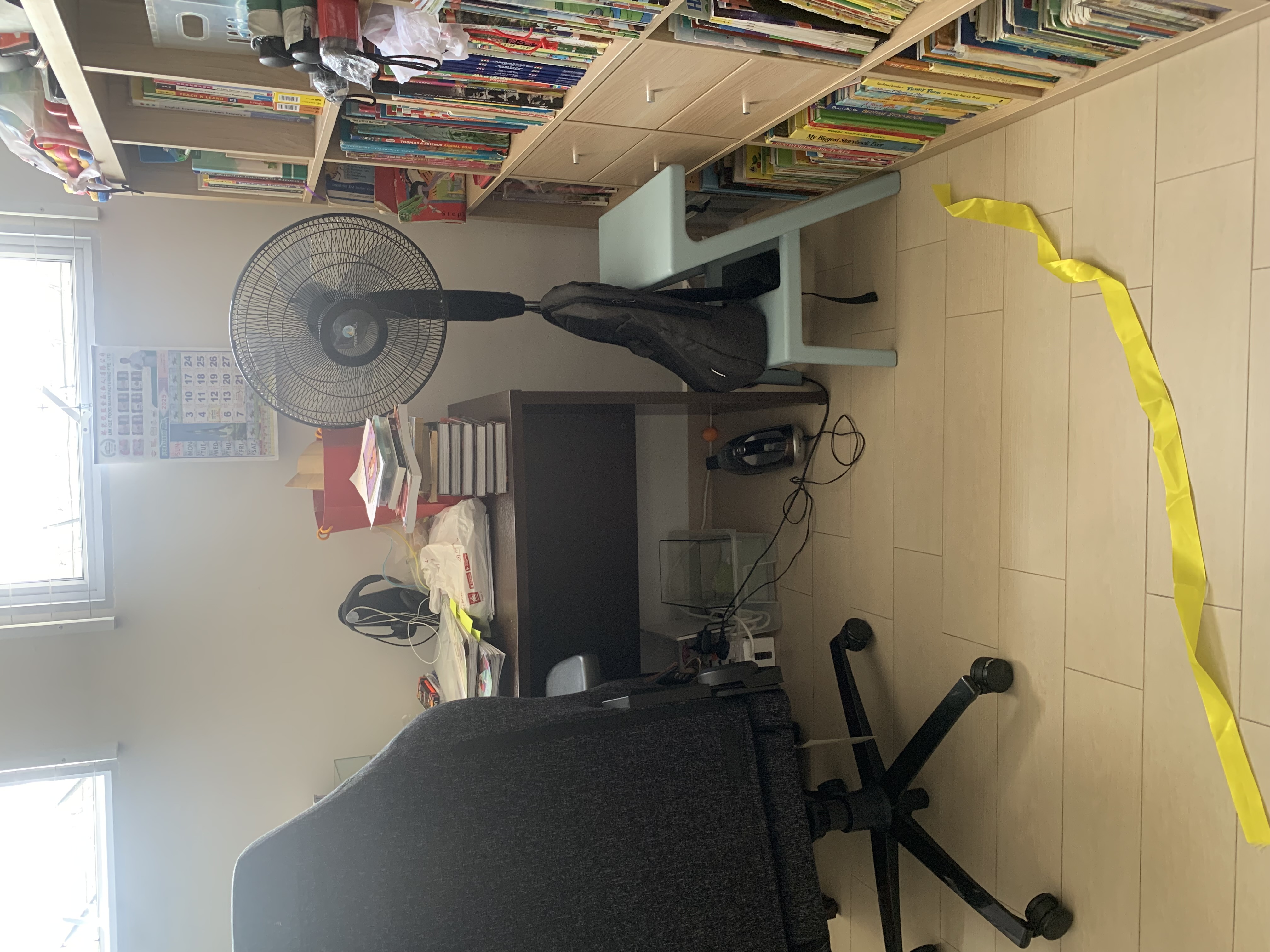}} \enskip
\subfloat[]{\includegraphics[angle=270, width = 0.3\linewidth]{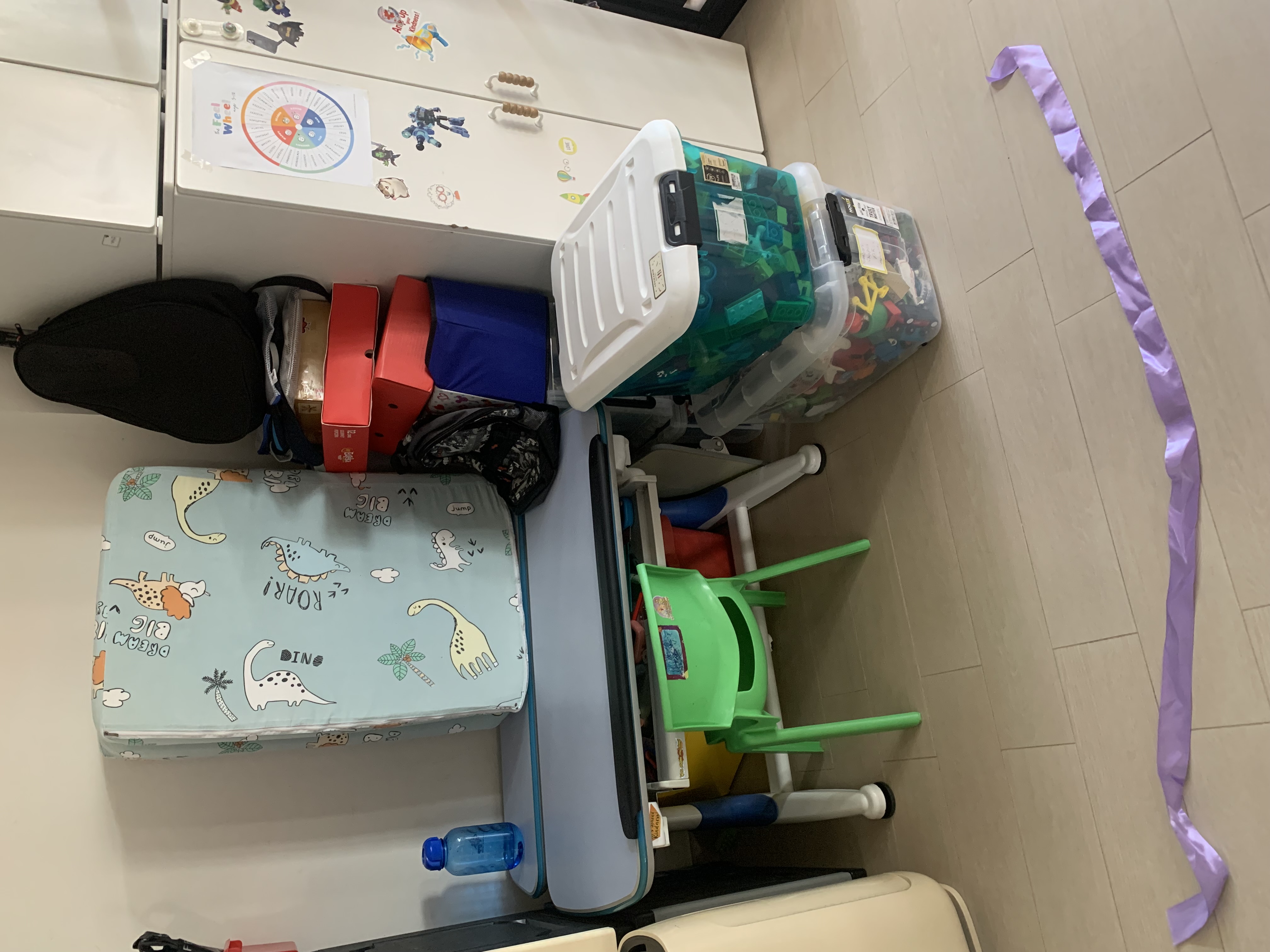}}
\caption{\change{Photos from P22’s probe activity (Activity 2) illustrating her attempts to carve out workspace at home. These images show how unequal power dynamics and unsuitable physical settings constrained her ability to work}: her husband’s room (left) was available only temporarily and under his conditions; the living room corner (center) became too messy and stuffy; and the children’s study (right) was physically uncomfortable.}
\Description{Three images labelled (a), (b), and (c), depicting spaces in P22's home. Image (a) shows P22's husabnd's bedroom with a workdesk and office chair on the left of the photo, a massage chari behind it, and a single bed on the right of the image. (b) shows a corner of P22's living room, with a desk placed against the wall, with an office chair. There is a standing fan to the right of the desk and a wall of shelves to the right of the image. Image (c) shows the children's study, with a small low table and chair that is sized for children. There are various items stacked in a corner of table and two large storage boxes to the right of the chair. }
\label{fig:p22home}
\end{figure*}

In another case, P22, a mother managing childcare and eldercare, \change{faced limited household support and no extended kin to lean on emotionally, relying instead on paid domestic workers solely for practical support in managing work and care (see Case Study 2 in Appendix A for more details). Unequal power dynamics and unsuitable physical settings further constrained her ability to work (see Figure~\ref{fig:p22home}). Without relational buffers, she turned inward to small spatial and material practices, such as pausing at the kitchen counter with tea or fragrance samples, to create brief moments of quiet.} While these practices could not eliminate stress, they provided small pauses of respite that helped her cope amidst daily demands. 


\subsubsection{Institutional Assets Supporting Flexibility} \label{finding-institutional-1}
Beyond the family, participants' maintenance of daily rhythms is also contingent on institutional and relational assets at the workplace. Supportive policies, empathetic supervisors, and collegial cultures often enabled participants to manage care alongside work. Some participants in large multinational firms, \change{such as P2 and P16} described supportive policies and cultures, \change{which} provided them considerable agency over their time with little pushback when care needs arose. P16 could reschedule meetings without justification, \inlinequote{I'll just say, sorry, I have a conflict.}{P16}, and colleagues were understanding even when care appointments overran. 
Similarly, policies and work culture that normalized work from anywhere enabled P2 to draw on institutional assets seamlessly \inlinequote{...everyone is okay not seeing you in the office because the travel in this company happens so frequently.}{P2}.

\begin{figure*}[ht]
 \centering
  \subfloat[]{\includegraphics[width = 0.45\linewidth]{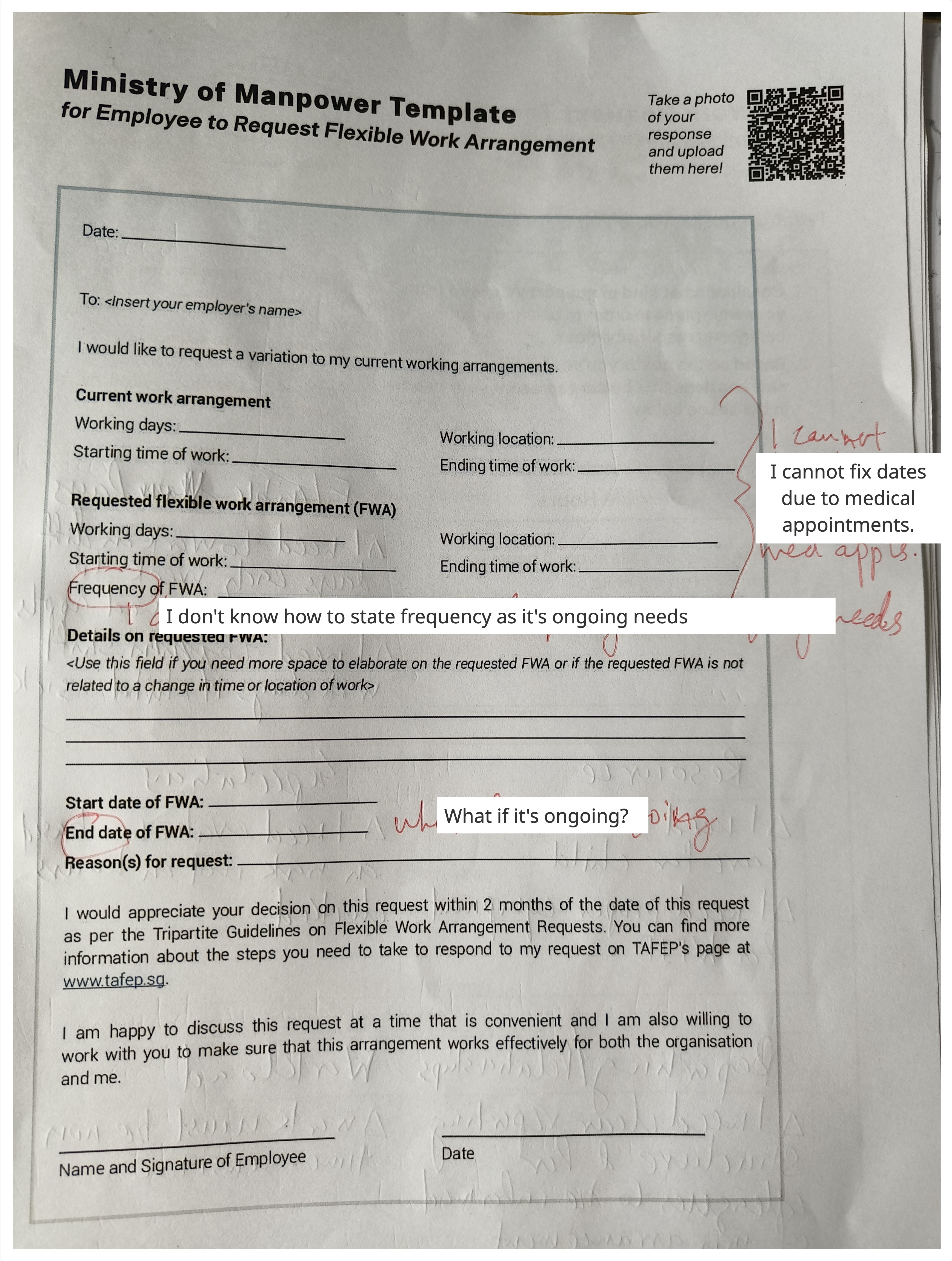}} \enskip
\subfloat[]{\includegraphics[width = 0.45\linewidth]{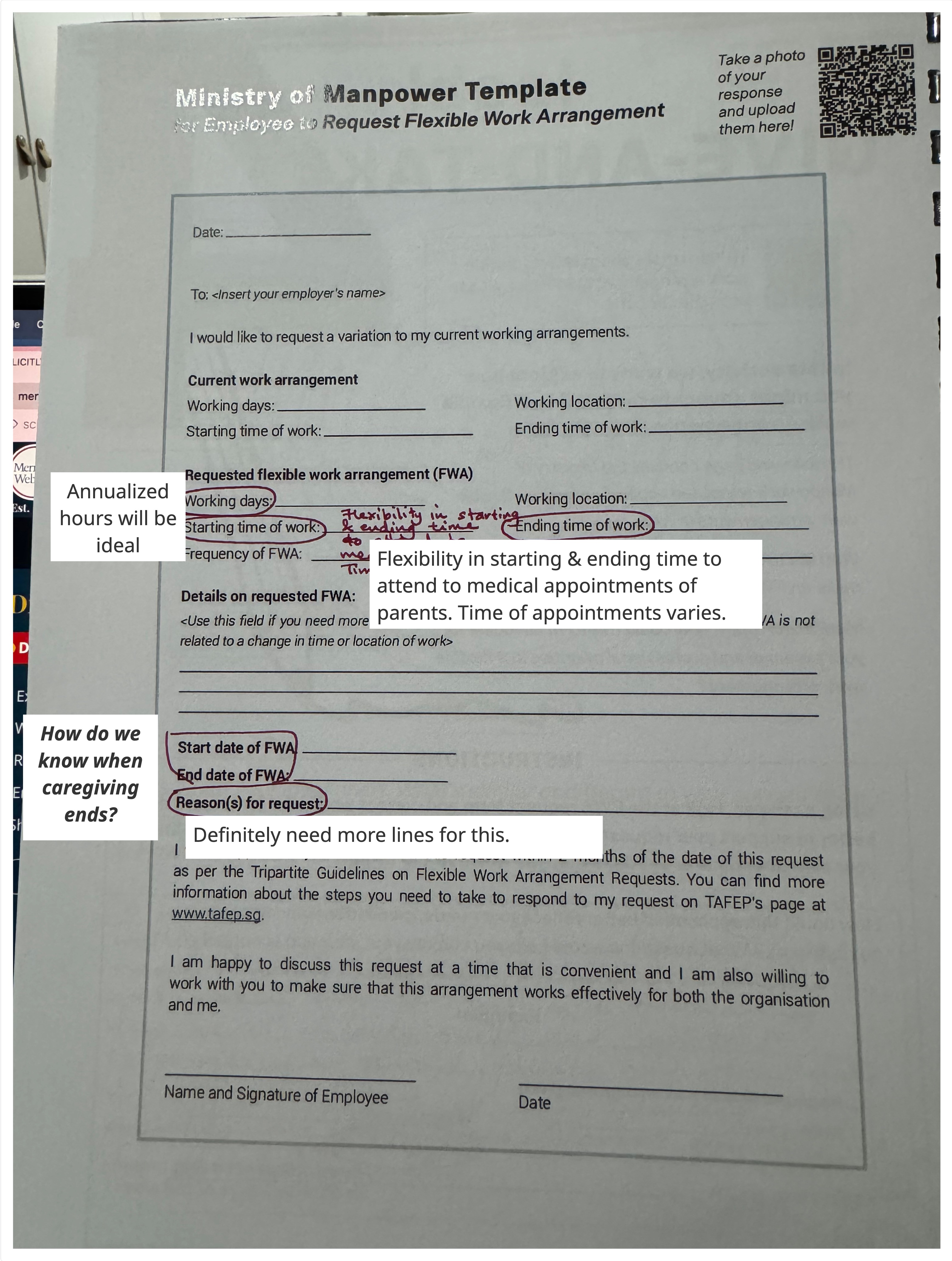}}
\caption{Participants’ responses to the Ministry of Manpower’s Flexible Work Arrangement request form (\change{design} probe activity 4): They described its rigid requirements for fixed times, frequencies, and end dates as misaligned with the ongoing and unpredictable nature of caregiving.  \inlinequote{I don’t know how to state frequency as its ongoing needs.}{P12} (left); and \inlinequote{How do we know when caregiving ends?}{P5} (right).}
\Description{Two images labelled (a) and (b) showing forms with fields such as 'current work arrangement', 'requested flexible work arrangement (FWA)', 'details on requested FWA', 'start and end dates for FWA' and 'reason(s) for request'. On the form, there are participants' written annotations in red.}
\label{fig:activity4}
\end{figure*}

\change{However, those in smaller companies encountered fewer formal accommodations,} limiting their ability to sustain workable rhythms. \change{While the national Flexible Work Arrangement request form applies across all workplaces, participants in such settings lacked internal policies or supportive infrastructures, and thus felt its limitations more acutely as the form’s rigid criteria did not reflect the fluid,} ongoing nature of caregiving (see Figure~\ref{fig:activity4} for more detail). In this context, \change{participants drew upon supportive workplace culture such as manager’s approval for time off, and colleague’s willingness to cover tasks in their absence to soften institutional gaps, requiring substantial relational work to negotiate and maintain.} For example, P8 \change{faces limitations in flexibility due to policy and structural constraints,} as her workplace requires adequate on-site staffing for each shift. \change{When care needs arise,} she draws on a trusting relationship with her supervisor to negotiate care leave. By requesting only when necessary, she builds credibility, making approval more likely, \inlinequote{...my boss knows that if you hear from (me), that means…really \textit{zóu tóu wú lù}\footnote{Pinyin romanization is used for readability} (she had no choice)...have to approve.}{P8}. Likewise, empathetic colleagues and a culture of mutual help enabled P19 and P5 to leave early or swap duties when care needs arose: \inlinequote{...as long as the work has someone to cover…Some of my colleagues are also caregiving. So all of us are like in the same boat…very understanding…we cover each other's stuff.}{P19}. 

\subsubsection{When Institutional Assets Falter} \label{finding-institutional-2}
\change{While participants leverage a supportive workplace culture to buffer policy gaps} such configurations remain informal and therefore precarious, particularly when relational dynamics within the workplace are tense or unsympathetic. In P12’s case, flexible work arrangement was possible only because her director \inlinequote{carved out the job}{P12} \change{for her.} However, most of her colleagues were younger, without caregiving responsibilities, and worked primarily from the office, shaping norms that made her flexible rhythms feel out of place. \change{To cope, she proactively manages colleagues’ perceptions to maintain the relational goodwill she relies on for flexibility---for example, attuning to their working hours when sending updates, reassuring them of her availability, and negotiating boundaries tactfully: } 

\begin{quote}
	\textit{“I also don't want to come across as being difficult. I try my best to complete… Sometimes subtly I will just remind them...it's actually on my off…the thing that you requested is actually not within my scope…”} (P12)
\end{quote}
\change{While these negotiations} sustained her arrangement, its legitimacy depended on her director’s presence: \inlinequote{...my director might be leaving so if I don't have somebody who…believes in me then it's very difficult.}{P12}. Without policy protections, flexibility can hinge on fragile workplace relationships.

Others described unsupportive supervisors and cultures that made formal policies feel hollow. In P24’s case, although workplace policies formally allowed flexibility, she recalled being told remote work was a “privilege”. A gossip culture further heightened pressures to conform: \inlinequote{...we have one colleague who sometimes not at his desk and people talk about it behind his back. So yeah, I feel a pressure to sit at my desk all the time, even when I've got nothing to do.}{P24}. She avoided requesting flexibility and instead relied on her cousin to accompany her mother to medical appointments. P7 likewise faced colleagues’ skepticism, \inlinequote{they think when we work from home, we are not work}{P7}, which sometimes forced her back to the office.

Other participants chose to reject institutional arrangements altogether and construct their own alternate infrastructures. In one notable case, P10 and her husband both chose to give up full-time employment, configuring their own system of part-time work, lifestyle adjustments, and tag-teaming to align rhythms with their caregiving priorities. Despite relinquishing her career, P10 expressed no regret, \inlinequote{I feel grateful that I'm able to do that because I guess not everybody has that option.}{P10}, adding that \inlinequote{my values is more based in like relationships, family…}{P10}. Her synergistic coordination with her spouse became her key asset to build a shared rhythm that aligned with their deeply held values, prioritizing their family over career.

These cases illustrate that flexibility is not a static condition, but rather it is actively configured by participants through ongoing relational coordinations and negotiations across home and institutional settings. These negotiations constitute an often invisible but essential infrastructure for sustaining participants’ rhythms, making flexibility possible. At the same time, organizational cultures, workplace policies, and household dynamics can either support these rhythms or, conversely, reinforce stigma, impose penalties or create additional burdens. In this way, relational and institutional assets---and the conditions under which they falter---fundamentally shape whether rhythms can be collectively sustained, or whether the burden falls disproportionately on individual participants.

\subsection{Sustaining Rhythms through Intrapersonal Assets}  \label{finding-rhythm-3}
While external supports shape the conditions of flexibility, they did not always fully buffer the strains of managing dual responsibilities. \change{Some participants} often described sacrificing aspects of themselves (i.e., their careers, health, personal time) to preserve workable rhythms for care and work. Even so, they found ways to sustain themselves amidst daily responsibilities by leveraging intrapersonal assets in the form of everyday self-care practices such as micro-boundary work, engaging in screen-based media, and cultivating self-awareness and positive inner orientations.

\subsubsection{Micro-Boundary Practices for Emotional Reset} \label{finding-micro}
Amidst daily demands, \change{participants used micro-boundary practices to momentarily reset their emotional state and regain capacity for work and care. P16 turned to beading as a brief pause between tasks, describing it as her } \inlinequote{own Zen moment in midst of all the chaos… So it's a way of just, like, mentally detaching for a while and getting a bit of peace before going back to whatever call or email.}{P16}. 
Similarly, P5 walked on the treadmill to release the stress of caregiving and return to work with renewed presence: \change{\inlinequote{I couldn't focus on my work… so I walk on the treadmill for about 30 minutes to get it out of my system before I go to class}{P5}.} These examples extend prior HCI research on micro-boundary work~\cite{cecchinato2015micro}, showing that workers rely not only on temporal strategies but also on affective and material anchors to reframe emotional states and sustain long hours across entangled care and work demands.



\subsubsection{Screen-Based Media as Everyday Coping Resource} \label{finding-screen}
Participants reported relying on \change{screen-based media} such as watching Netflix or scrolling through social media, \change{to gain reprieve from daily responsibilities}. Even as they are aware that these activities were not restorative \change{and \inlinequote{mindless}{P6}}, \change{they engaged it as a form of low-effort, “battery conservation mode” of rest.} As P6 describes it, \inlinequote{I really didn't have energy... my rest was not active rest, because actually active rest…It requires effort to a certain extent, so it was more, like, mindless stuff, like Netflix…so that I can continue conserving my battery, and then be available to resolve any emergency that comes up.}{P6}. \change{Similarly, others like P16 treated it as a form of quick escape,} \inlinequote{sometimes I watch Netflix because it's a bit mindless…help me to get away from, you know.}{P16}. 

\change{Yet others, like P9, used social media for reflection and to stay connected to community} (\inlinequote{...on social media...people who are in similar, under some similar circumstances, um, share what they do or their day-to-day actually does, you know, make a difference in, in supporting my wellbeing.}{P9}). \change{Another participant, P1, used it} as a private outlet for reflection on daily life, which she found particularly supportive as a single parent (see Appendix~\ref{case} for details).  
The use of screen-based media for self-care is often criticized as detrimental. \change{Though imperfect,} our study reveals a more nuanced role–for some, \change{it functions as a flexible and meaningful tool for self care. These low-effort forms of digital engagement offered overburdened participants immediate mental relief when more restorative practices required energy they did not have.}

\subsubsection{Cultivating \change{Helpful} Inner Orientations} \label{finding-inner}


Participants also cultivated inner orientations that helped them cope with the ongoing demands of caregiving. Rather than trying to eliminate stress---which they recognized as inevitable---they adapted how they related to it through reframing and acceptance. For instance, P16 described meeting the messiness of blurred boundaries with a “live in the moment” approach: \inlinequote{I try to live in the moment… If I need to work at 8pm, I don't get too upset. I just get it done}{P16}. P10 similarly redefined what counted as “enough,” \change{focusing on affection, presence, and basic needs (\inlinequote{just enough food… just enough fun… surviving}{P10}).}

\change{Participants also emphasized self-awareness as a way to protect their emotional energy. P9 described tuning into bodily signals to avoid burnout: \inlinequote{If I'm getting overwhelmed or very tired… that means I'm reaching burnout… and that's something I want to avoid}{P9}.}

Others emphasized cultivating gratitude and self-compassion. P9 shared the importance of starting her day “right” by appreciating small things, \inlinequote{ I think that's so important for me, especially because I really do lead a challenging life…it can be really highly stressful.}{P9}. Likewise, P13 stressed, \inlinequote{End of day, I just want to say we have to be kind to ourselves...  So always give some leeway to ourselves.}{P13}. At the core of their practices is an orientation towards acceptance and care---for their families and themselves---enabling them to sustain the ongoing demands of work and caregiving.

Participants' intrapersonal assets are crucial for sustaining rhythms, even when the cost is sometimes self-sacrifice. Their layered practices and strategies reveal how they negotiate between care, work and self to uphold a liveable balance through inner resilience rather than solely through external support.

\section{Discussion} 


Our study extends understanding \change{on how flexible workers in caregiving contexts} \change{make rhythms of work and care by articulating and negotiating temporal, spatial and relational assets to manage the entanglement of both roles}. \change{In this section, we first discuss the design implications of reframing blurred boundaries as assets rather than deficits. We then examine the critical role of human infrastructure in sustaining these rhythms. Next, we examine how inequitable access to such supports shapes flexible workers’ experiences. We conclude with the study’s limitations and directions for future research.}


\subsection{\change{A Time Well Spent Perspective for} Designing for Flexible Workers}
\change{Previous work on HCI for supporting WLB has proposed tools to delineate clear temporal boundaries, managing task allocation based on productivity and efficiency principles~\cite{guillou2020your, saha2023time, das2023focused, pschetz2015time}. However,} our findings show that instead of structured time management, \change{participants} adopt strategies that leverage temporal and spatial blurred boundaries to \change{maintain interleaved work-care rhythms, allowing them to alternate between roles in response to contingencies (e.g., Section~\ref{finding-rhythm-1}).} Building on these insights, we turn to discuss how technologies could support flexibility as it is lived in practice, shaped by the ongoing negotiation of everyday contingencies. 



\change{Participants’ appreciation for blurred boundaries, aligns with existing literature describing more fluid engagement with temporal rhythms when flexible work is negotiated with caregiving responsibilities, valuing flexibility to respond to the spontaneity of life events~\cite{erickson2019flexible, steup2022attune}}. Our findings showed that much of \change{their} rhythm-making relied on invisible articulation \change{of temporal and spatial assets, as well as} intrapersonal \change{assets, to maintain and sustain interleaved works and care rhythms.} \change{These strategies were accompanied by} challenges such as emotional and financial strain, domestic role management, extended working hours, and heightened expectations, as reported in prior research~\cite{duxbury2009balancing, hsu2024dancing, kabir2025balancing}, surfacing the trade-offs participants were willing to make (e.g., Section~\ref{finding-coarse}). Yet, \change{despite being central,} these \change{articulation} practices \change{are} rarely \change{recognized} in conventional productivity systems, which privilege visible, measurable outputs.

To address this, a \emph{Time Well Spent} perspective~\cite{chow2023feeling, guillou2020your}, which shifts the frame towards valuing \change{beyond waged labour tasks}, can be used to surface \change{individuals' strategies for sustaining work-care rhythms, as part of assessing a successful day}. However, just making these practices visible is not enough; designing for flexible work should avoid monitoring or optimizing these practices and instead affirm their legitimacy \change{and revalue its role} as part of a productive day. For example, tools could highlight \change{the efforts of rescheduling tasks and reconfiguring daily calendars as part of the work performed in a day to sustain a livable life}. \change{Moreover, individuals' } self-care moments \change{could also be legitimized as an integral component of broadening perspectives of} what constitutes a “successful day”~\cite{avrahami2020celebrating}.  

\change{The development of these support systems, require in turn a reframing of the temporal dimension.} Participants’ experiences highlight that time was lived as \change{fluid} and interwoven, not as neatly segmented units. Existing task and time management systems, grounded in the logic of \emph{circumscribed time}~\cite{mazmanian2015circumscribed}, \change{do not fulfill individuals' needs of fluidity, as expressed by P20 (e.g., Section~\ref{finding-coarse})}. \change{Thereby, we argue the need to move away from logics of chunkable segmentation of time~\cite{pschetz2015time}, towards a more situated approach that captures the tacit expertise of individuals as they articulate their work and care rhythms in negotiation with external biological, or organizational rhythms~\cite{leshed2014farm}.} \change{Reframing time through approaches such as mosaic like temporalities~\cite{mazmanian2015circumscribed}} could allow \change{flexible} workers to layer overlapping activities in ways that mirror their everyday negotiations. Rather than functioning solely as schedulers or optimizers, technologies could help visualize and reflect on \change{their fluid rhythms}, foregrounding ongoing \change{maintenance and} negotiation as central to daily life. 

While such fluidity is difficult to render with conventional calendars or timelines, a few studies have begun to explore alternatives (e.g.,~\cite{lupi2016dear, brock2024sonificaiton, thudt2018reflection}). For example, Snyder et al.~\cite{snyder2019bipolar} explore artistic and experiential encodings that foreground lived complexity rather than discrete events, showing how temporal representations can surface subjective, embodied dimensions of experience. \change{In this line, participants' practices such as self-awareness, mindless rest, and perspectives of ``lived in the moment'' could be revalued through  \emph{somadata} approaches~\cite{alfaras2020somadata}, that have demonstrated how rhythms can be \emph{felt} as well as seen by coupling biodata to sound, heat, or movement, opening multimodal representations that resonate with bodily cadence (e.g. \cite{strahl2022making}).} Such recognition not only supports individuals’ well-being but also counters deficit framings of flexible work as fragmented or unproductive. While broader structural issues cannot be solved through technology alone, amplifying intrapersonal resilience practices represents a crucial step toward more \change{tailored support systems}.

\subsection{Reframing Infrastructure for Flexible Work}

A crucial aspect highlighted in our findings is the role of  relational and organizational assets \change{in supporting how flexible workers with caregiving responsibilities sustain their daily rhythms}. Our findings showed that individuals with stronger organizational support could engage more proactively in managing rhythms of care and work. However, institutional assets, such as well-organized company policies or supportive culture alone do not guarantee better WLB. Our findings pointed out that support from family, extended kin, and close networks is also essential for sustaining work–care rhythms (e.g., Section~\ref{finding-relational-1}). In this context, social support played a vital role in creating flexible work routines, serving as a form of \emph{human infrastructure}~\cite{lee2006human} that filled gaps when national, organizational, and individual resources fell short. For instance, participants coordinated long-term plans, weekly schedules, and negotiated  flexibility with household co-caregivers to establish mutual rhythms, drawing on both their own assets and those of their co-caregivers’. This extension of organizational resources into the family sphere illustrates how such support benefits not only employees but also their family members, who may lack equivalent organizational backing~\cite{voydanoff2004effects}. 

These findings extend the literature on flexible work infrastructure  which has primarily examined the work sphere by focusing information technologies role in enabling flexibility and managing boundaries~\cite{jarrahi2022digital, pei2019we, thomson2013information, salazar2001building}. \change{In particular, our findings shed light on how flexibility is negotiated with others, when caregiving tasks are a central part of one's life.} While these approaches aim at reducing work–life conflict and facilitating collaboration between co-workers, less attention has been given to the role of \emph{human infrastructure}, particularly the relational supports within the home \change{to distribute caregiving responsibilities}. Our findings highlight a crucial role of relational assets \change{for effective family-supportive environments} interwoven in flexible work infrastructure, extending  frameworks like the one outlined by~\citet{kossek2025reenvisioning}, which delineates national, organizational and individual levels of effective policies. 

The case of P22 illustrates these dynamics in practice, highlighting the challenges that arise when \change{household relational assets are} limited. Although her institutional assets provided autonomy at work, the absence of family support constrained her ability to maintain stable work-care rhythms. Facing a power imbalance in access to time and space, she relied heavily on her intrapersonal and paid human infrastructure resources to navigate work and care (Section~\ref{finding-relational-2}). Her experience underscores the critical role of domestic workers as a foundational layer of human infrastructure for sustaining daily rhythms~\cite{teo2018whose}. Yet, leveraging this additional workforce is not without cost, it introduces further coordination demands to the main caregiver~\cite{hsu2024dancing, kabir2025balancing}, requiring careful negotiation of schedules, responsibilities, and shared expectations. Her experiences highlight the extent to which sustaining rhythms requires not only organizational and material assets, but also the relational supports that constitute human infrastructure.

Relational \change{assest are crucial for enabling} \change{flexible workers with care responsibilities} to redistribute care burdens and maintain workable rhythms amid daily stresses. These insights suggest the need for research on flexible work infrastructure \change{that explores beyond the individual worker to}  account for the relational dynamics that shape individuals’ experiences. This shift raises questions like: How might technologies be developed to recognize and strengthen the relational assets supporting worker’s flexible arrangements? In organizational settings where supportive cultures are lacking, could design interventions open pathways for more equitable and flexible collaboration? These are provocations to invite future research to reimagine infrastructural support for flexible work not as an individual productivity problem or solely an organizational concern, but as an ecology of human, organizational, and digital arrangements sustained through everyday negotiations. 


\subsection{Future Directions for More Just Flexible Work Arrangements}

Public policies in Singapore have promoted flexible work arrangements through initiatives intended to help organizations support employees with caregiving responsibilities. However, our study reveals that the level of support remains uneven: only \change{participants working} in MNCs benefited from robust policies and a supportive culture. Others encountered limited flexibility, often facing trade-offs such as shifting from full-time to part-time work to gain autonomy or sacrificing career advancement to meet caregiving demands. Prior research has shown that the work–care policy landscape largely supports higher-income families, reinforces nuclear family assumptions, and frames care needs as individualized~\cite{teo2018whose}. Building on our findings and these critiques, we propose that a design justice~\cite{costanza2020design} perspective could help extend the benefits of flexible work to those navigating more complex caregiving contexts.

National instruments aimed at improving workers' WLB through \change{FWAs} \change{might} fail to accommodate actual lived experiences (e.g., Fig.~\ref{fig:activity4}). Dominant perspectives for WLB are based on the idea of segmenting time and space to increase productivity. In our study, participants observed that this approach does not accurately reflect their work and care demands, where responsibilities are not scheduled in a consistent or linear manner but are instead dynamically assessed. \change{Therefore, }these instruments generate what \citet {kossek2021future} describes as \emph{inflexible flexibility}, where instead of assisting workers \change{equally}, they \change{produce an uneven landscape where those with fewer assets to manage and maintain work rhythms end up} making trade-offs in other areas of their lives.

Rooted in a design justice approach, foregrounding who participates in design, how it reproduces inequalities, and whose experiences are recognized, we argue that \emph{true flexibility}~\cite{kossek2021future} must be defined with attention to the diverse circumstances individuals face. To guide policy development for true flexibility, research should draw on co-design and participatory methods with individuals navigating complex situations, such as limited family or organizational support, high caregiving demands, or financial hardship. These approaches could help prevent policies from reproducing inequality while supporting individuals’ rhythm-making practices. 



\subsection{Limitations and Future Work}

\change{Reflecting on the transferability of our findings, we note that our insights are shaped by the specificities of the Singapore context. As a highly developed economy with strong collectivist values, Singapore represents a unique intersection of Global North infrastructures and family-centric cultural norms. Consequently, our insights regarding the reliance on extended kin and domestic help are most directly transferable to contexts with similar family-centric support structures, rather than those with purely individualistic or state-welfare models. Nevertheless, the broader insight (e.g., blurred temporal boundaries as a resource, and recognizing human and organizational infrastructures as essential assets) remains relevant across diverse settings~\cite{ciolfi2018work, kabir2025balancing}. Future studies could deepen this comparative perspective by examining how flexible workers in different cultural and policy environments mobilize (or lack access to) these infrastructures, and how such variations shape their capacity to enact livable work–care rhythms.}

\change{We also caution against over-generalizing our findings to all flexible workers. Our study focused on individuals who actively use flexible work to meet caregiving responsibilities. Prior work shows that workers with greater spatial and temporal agency (e.g., single adults) often prefer segmented and office-like strategies for structuring flexibility~\cite{cho2022topophilia}. In contrast, our participants, who lack this full agency due to care demands, must negotiate and complement their constraints through their own unique rhythm making. Although our findings do not represent all flexible workers, we argue that this work presents a novel and necessary perspective on the infrastructure required for those engaged in the double burden of work and care.}

Also, our study engaged a broad and diverse pool of participants with varying work, care and household contexts. While our diverse sampling enabled a broad understanding of how flexible workers navigated work-care arrangements, our analysis also revealed the highly situated \change{and gendered} nature of these practices. \change{ Our participants were predominantly female, reflecting realities where caregiving responsibilities disproportionately fall on women~\cite{chung2020flexible}. While we did not explicitly screen for gender, this distribution indicates that our findings regarding their rhythm-making are deeply informed by the female experiences. Beyond gender} each participant’s rhythms and access to assets were shaped by intersecting factors–including caregiving type, household structure, employment status, and socio-economic positioning. For instance, a young, only child caring for an elderly parent may face limited relational assets within the household, and heightened vulnerability to supervisory bias, especially if employed in junior roles with limited institutional protections. In contrast, workers in upper management or dual-income households may have greater leverage in negotiating flexibility or accessing paid support such as live-in domestic help. Immigrant families may lack extended kin networks, while parents of young children face different temporal demands than those caring for elderly relatives. While we cannot exhaustively map all such configurations, these nuances point to the need for deeper, population-specific inquiry. Notably, our sample included several single parents (N=4) and parent(s) of children with special needs (N=4). While this may reflect snowball sampling dynamics, it nevertheless highlights underrepresented groups navigating caregiving systems not designed for their needs. In the Singapore context, single parents often face social and policy stigma~\cite{wong2004spaces}, while parents of children with special needs bear greater care and financial burdens. Examining how particular structural positions and caregiving contexts shape access to flexibility is crucial to offer more targeted insights for equitable policy and design.

Our methods provided valuable insights  into how domestic spaces were arranged to support both work and care. However, because they offered only partial and mediated views of home life, they could not fully convey the nuances of everyday spatial practices. We had initially aimed to conduct interviews in participants’ homes to better capture the spatial dimensions of their work-care rhythm configurations. However, all but one participant opted for online interviews over the home visit. Future research could explore this more deeply through longer-term in-home ethnography.

Lastly, we found in our study that live-in domestic workers play a critical role in supporting caregiving and household continuity, influencing participants’ capacity to enact flexibility. While our study captured aspects of participants’ coordination with live-in domestic workers, it did not explicitly address the power dynamics within these relationships, given that data was drawn solely from interviews with flexible workers who were their employers. This opens up opportunities for future research to attend to the perspectives of live-in domestic workers themselves, who remain underrepresented despite being central to Singapore's care infrastructure.
 
\section{Conclusion}
Individuals configure flexible arrangements within diverse work and home conditions. While earlier studies examined work-life balance and flexible work arrangements primarily from productivity and organizational perspectives, our study foregrounds its intersection with workers' caregiving roles and domestic life. Using an asset-based lens, we show how workers sustain work-care rhythms by leveraging temporal, spatial, relational, institutional, and intrapersonal assets–even amidst structural constraints. Our findings reframe blurred boundaries not as failures of boundary management, but as rhythms to be actively sustained. They also highlight flexible rhythms as collectively sustained rather than individually configured, and therefore shaped by structural conditions. This calls for alternative design approaches that embrace the messiness of flexibility, strengthen workers' relational assets, and critically attend to workers' varied structural conditions. To more deeply understand the rhythm-making strategies of diverse worker-caregiver conditions, we invite future research focused on specific subpopulations, particularly those underrepresented in existing scholarship. We hope that these insights support the development of tools and systems that are more attuned to the situated realities of flexible workers managing care.

\begin{acks}
We thank the Associate Chairs and reviewers for their thoughtful feedback, as well as members of the Joyful Experiences in Design and Interaction (JEDI) Lab for their valuable input. This work was supported by the MOE Academic Research Fund (AcRF) Tier 1 FRC Research Grant, Ministry of Education, Singapore (ID: A-8003205-00-00). This research is dedicated to those who navigate the intensity of everyday life at work while also caring for the people they love, and we are especially grateful to the participants who shared their stories with us.
\end{acks}

\bibliographystyle{ACM-Reference-Format}
\bibliography{CHI2026.bib}

@inproceedings{hsu2024dancing,
author = {Hsu, Long-Jing and Chung, Chia-Fang},
title = {Dancing with the Roles: Towards Designing Technology that Supports the Multifaceted Roles of Caregivers for Older Adults},
year = {2024},
isbn = {9798400703300},
publisher = {Association for Computing Machinery},
address = {New York, NY, USA},
url = {https://doi.org/10.1145/3613904.3642728},
doi = {10.1145/3613904.3642728},
booktitle = {Proceedings of the 2024 CHI Conference on Human Factors in Computing Systems},
articleno = {1010},
numpages = {12},
location = {Honolulu, HI, USA},
series = {CHI '24}
}

@inproceedings{leshed2014farm,
author = {Leshed, Gilly and H\r{a}kansson, Maria and Kaye, Joseph 'Jofish'},
title = {"Our life is the farm and farming is our life": home-work coordination in organic farm families},
year = {2014},
isbn = {9781450325400},
publisher = {Association for Computing Machinery},
address = {New York, NY, USA},
url = {https://doi.org/10.1145/2531602.2531708},
doi = {10.1145/2531602.2531708},
booktitle = {Proceedings of the 17th ACM Conference on Computer Supported Cooperative Work \& Social Computing},
pages = {487–498},
numpages = {12},
location = {Baltimore, Maryland, USA},
series = {CSCW '14}
}

@article{kabir2025balancing,
author = {Kabir, Kazi Sinthia and Cohoon, Johanna and Lund, John R. and Chen, Jacqueline M. and Metz, A.J. and Wiese, Jason},
title = {Balancing Care Responsibilities with Remote Work},
year = {2025},
issue_date = {January 2025},
publisher = {Association for Computing Machinery},
address = {New York, NY, USA},
volume = {9},
number = {1},
url = {https://doi.org/10.1145/3701182},
doi = {10.1145/3701182},
journal = {Proc. ACM Hum.-Comput. Interact.},
month = jan,
articleno = {GROUP3},
numpages = {21}
}

@article{sefidgar2024nonwork,
author = {Sefidgar, Yasaman S. and J\"{o}rke, Matthew and Suh, Jina and Saha, Koustuv and Iqbal, Shamsi and Ramos, Gonzalo and Czerwinski, Mary},
title = {Improving Work-Nonwork Balance with Data-Driven Implementation Intention and Mental Contrasting},
year = {2024},
issue_date = {April 2024},
publisher = {Association for Computing Machinery},
address = {New York, NY, USA},
volume = {8},
number = {CSCW1},
url = {https://doi.org/10.1145/3637351},
doi = {10.1145/3637351},
journal = {Proc. ACM Hum.-Comput. Interact.},
month = apr,
articleno = {74},
numpages = {29}
}

@article{erickson2019flexible,
  title={Flexible turtles and elastic octopi: Exploring agile practice in knowledge work},
  author={Erickson, Ingrid and Menezes, Deepti and Raheja, Raghav and Shetty, Thanushree},
  journal={Computer Supported Cooperative Work (CSCW)},
  volume={28},
  number={3},
  pages={627--653},
  year={2019},
  publisher={Springer}
}

@article{ciolfi2018work,
  title={From Work to Life and back again: Examining the digitally-mediated work/life practices of a group of knowledge workers},
  author={Ciolfi, Luigina and Lockley, Eleanor},
  journal={Computer Supported Cooperative Work (CSCW)},
  volume={27},
  number={3},
  pages={803--839},
  year={2018},
  publisher={Springer}
}

@article{cho2022topophilia,
author = {Cho, Janghee and Beck, Samuel and Voida, Stephen},
title = {Topophilia, Placemaking, and Boundary Work: Exploring the Psycho-Social Impact of the COVID-19 Work-From-Home Experience},
year = {2022},
issue_date = {January 2022},
publisher = {Association for Computing Machinery},
address = {New York, NY, USA},
volume = {6},
number = {GROUP},
url = {https://doi.org/10.1145/3492843},
doi = {10.1145/3492843},
journal = {Proc. ACM Hum.-Comput. Interact.},
month = jan,
articleno = {24},
numpages = {33}
}

@article{breideband2022teamwork,
author = {Breideband, Thomas and Talkad Sukumar, Poorna and Mark, Gloria and Caruso, Megan and D'Mello, Sidney and Striegel, Aaron D.},
title = {Home-Life and Work Rhythm Diversity in Distributed Teamwork: A Study with Information Workers during the COVID-19 Pandemic},
year = {2022},
issue_date = {April 2022},
publisher = {Association for Computing Machinery},
address = {New York, NY, USA},
volume = {6},
number = {CSCW1},
url = {https://doi.org/10.1145/3512942},
doi = {10.1145/3512942},
journal = {Proc. ACM Hum.-Comput. Interact.},
month = apr,
articleno = {95},
numpages = {23}
}

@book{mattelmaki2006design,
  title={Design probes},
  author={Mattelm{\"a}ki, Tuuli},
  year={2006},
  publisher={Aalto University},
  address = {Vaajakoski, Finland},
}

@article{gaver1999probes,
author = {Gaver, Bill and Dunne, Tony and Pacenti, Elena},
title = {Design: Cultural probes},
year = {1999},
issue_date = {Jan./Feb. 1999},
publisher = {Association for Computing Machinery},
address = {New York, NY, USA},
volume = {6},
number = {1},
issn = {1072-5520},
url = {https://doi-org.libproxy1.nus.edu.sg/10.1145/291224.291235},
doi = {10.1145/291224.291235},
journal = {Interactions},
month = jan,
pages = {21–29},
numpages = {9}
}

@inproceedings{wallace2013making,
author = {Wallace, Jayne and McCarthy, John and Wright, Peter C. and Olivier, Patrick},
title = {Making design probes work},
year = {2013},
isbn = {9781450318990},
publisher = {Association for Computing Machinery},
address = {New York, NY, USA},
url = {https://doi.org/10.1145/2470654.2466473},
doi = {10.1145/2470654.2466473},
booktitle = {Proceedings of the SIGCHI Conference on Human Factors in Computing Systems},
pages = {3441–3450},
numpages = {10},
location = {Paris, France},
series = {CHI '13}
}

@inproceedings{boehner2007intepret,
author = {Boehner, Kirsten and Vertesi, Janet and Sengers, Phoebe and Dourish, Paul},
title = {How HCI interprets the probes},
year = {2007},
isbn = {9781595935939},
publisher = {Association for Computing Machinery},
address = {New York, NY, USA},
url = {https://doi.org/10.1145/1240624.1240789},
doi = {10.1145/1240624.1240789},
booktitle = {Proceedings of the SIGCHI Conference on Human Factors in Computing Systems},
pages = {1077–1086},
numpages = {10},
location = {San Jose, California, USA},
series = {CHI '07}
}

@article{charmaz2014constructing,
  title={Constructing grounded theory},
  author={Charmaz, Kathy},
  year={2014},
  publisher={SAGE publications Ltd}
}

@article{clarke2003situational,
  title={Situational analyses: Grounded theory mapping after the postmodern turn},
  author={Clarke, Adele E},
  journal={Symbolic interaction},
  volume={26},
  number={4},
  pages={553--576},
  year={2003},
  publisher={Wiley Online Library}
}

@inproceedings{wong2020assets,
author = {Wong-Villacres, Marisol and Gautam, Aakash and Roldan, Wendy and Pei, Lucy and Dickinson, Jessa and Ismail, Azra and DiSalvo, Betsy and Kumar, Neha and Clegg, Tammy and Erete, Sheena and Roden, Emily and Sambasivan, Nithya and Yip, Jason},
title = {From Needs to Strengths: Operationalizing an Assets-Based Design of Technology},
year = {2020},
isbn = {9781450380591},
publisher = {Association for Computing Machinery},
address = {New York, NY, USA},
url = {https://doi.org/10.1145/3406865.3418594},
doi = {10.1145/3406865.3418594},
booktitle = {Companion Publication of the 2020 Conference on Computer Supported Cooperative Work and Social Computing},
pages = {527–535},
numpages = {9},
location = {Virtual Event, USA},
series = {CSCW '20 Companion}
}

@article{shin2021cloud,
author = {Shin, Ji Youn and Chaar, Dima and Davis, Catherine and Choi, Sung Won and Lee, Hee Rin},
title = {Every Cloud Has a Silver Lining: Exploring Experiential Knowledge and Assets of Family Caregivers},
year = {2021},
issue_date = {October 2021},
publisher = {Association for Computing Machinery},
address = {New York, NY, USA},
volume = {5},
number = {CSCW2},
url = {https://doi.org/10.1145/3479560},
doi = {10.1145/3479560},
journal = {Proc. ACM Hum.-Comput. Interact.},
month = oct,
articleno = {416},
numpages = {25}
}

@book{saldana2015coding,
	Author = {Salda{\~n}a, Johnny},
	Publisher = {SAGE Publications},
	Title = {The coding manual for qualitative researchers},
	edition={3rd},
	address={Los Angeles, CA},
	Year = {2015}}

@techreport{teevan2022microsoft,
author = {Teevan, Jaime and Baym, Nancy and Butler, Jenna and Hecht, Brent and Jaffe, Sonia and Nowak, Kate and Sellen, Abigail and Yang, Longqi and Ash, Marcus and Awori, Kagonya and Bruch, Mia and Choudhury, Piali and Coleman, Adam and Counts, Scott and Cupala, Shiraz and Czerwinski, Mary and Doran, Ed and Fetterolf, Elizabeth and Gonzalez Franco, Mar and Gupta, Kunal and Halfaker, Aaron L and Hadley, Constance and Houck, Brian and Inkpen, Kori and Iqbal, Shamsi and Knudsen, Eric and Levine, Stacey and Lindley, Siân and Neville, Jennifer and O'Neill, Jacki and Pollak, Rick and Poznanski, Victor and Rintel, Sean and Shah, Neha Parikh and Suri, Siddharth and Troy, Adam D. and Wan, Mengting},
title = {Microsoft New Future of Work Report 2022},
institution = {Microsoft},
year = {2022},
month = {May},
url = {https://www.microsoft.com/en-us/research/publication/microsoft-new-future-of-work-report-2022/},
number = {MSR-TR-2022-3},
}

@inproceedings{fleck2015balancing,
author = {Fleck, Rowanne and Cox, Anna L. and Robison, Rosalyn A.V.},
title = {Balancing Boundaries: Using Multiple Devices to Manage Work-Life Balance},
year = {2015},
isbn = {9781450331456},
publisher = {Association for Computing Machinery},
address = {New York, NY, USA},
url = {https://doi-org.libproxy1.nus.edu.sg/10.1145/2702123.2702386},
doi = {10.1145/2702123.2702386},
booktitle = {Proceedings of the 33rd Annual ACM Conference on Human Factors in Computing Systems},
pages = {3985–3988},
numpages = {4},
location = {Seoul, Republic of Korea},
series = {CHI '15}
}

@article{semaan2019routine,
author = {Semaan, Bryan},
title = { 'Routine Infrastructuring' as 'Building Everyday Resilience with Technology': When Disruption Becomes Ordinary},
year = {2019},
issue_date = {November 2019},
publisher = {Association for Computing Machinery},
address = {New York, NY, USA},
volume = {3},
number = {CSCW},
url = {https://doi.org/10.1145/3359175},
doi = {10.1145/3359175},
journal = {Proc. ACM Hum.-Comput. Interact.},
month = nov,
articleno = {73},
numpages = {24}
}

@inproceedings{salazar2001building,
author = {Salazar, Christine},
title = {Building boundaries and negotiating work at home},
year = {2001},
isbn = {1581132948},
publisher = {Association for Computing Machinery},
address = {New York, NY, USA},
url = {https://doi.org/10.1145/500286.500311},
doi = {10.1145/500286.500311},
booktitle = {Proceedings of the 2001 ACM International Conference on Supporting Group Work},
pages = {162–170},
numpages = {9},
location = {Boulder, Colorado, USA},
series = {GROUP '01}
}

@article{kossek1999bridging,
  title={Bridging the work-family policy and productivity gap: A literature review},
  author={Kossek, Ellen Ernst and Ozeki, Cynthia},
  journal={Community, Work \& Family},
  volume={2},
  number={1},
  pages={7--32},
  year={1999},
  publisher={Taylor \& Francis}
}

@misc{white2010flexible,
  author       = "Cecilia Rouse",
  title        = "The Economics of Workplace Flexibility",
  howpublished = "https://obamawhitehouse.archives.gov/blog/2010/03/31/economics-workplace-flexibility",
  month        = "03",
  year         = "2010",
  note         = "Accessed: 2025-08-26",
}

@misc{tal2024fwa,
  author       = "Tripartite Alliance for Fair and Progressive Employment Practices",
  title        = "Tripartite Guidelines on Flexible Work Arrangement Requests",
  howpublished = "https://www.mom.gov.sg/-/media/mom/documents/press-releases/2024/tripartite-guidelines-on-flexible-work-arrangement-requests.pdf",
  month        = "04",
  year         = "2024",
  note         = "Accessed: 2025-08-26",
}

@article{gajendran2007good,
  title={The good, the bad, and the unknown about telecommuting: meta-analysis of psychological mediators and individual consequences.},
  author={Gajendran, Ravi S and Harrison, David A},
  journal={Journal of applied psychology},
  volume={92},
  number={6},
  pages={1524},
  year={2007},
  publisher={American Psychological Association}
}

@article{bailey2002review,
  title={A review of telework research: Findings, new directions, and lessons for the study of modern work},
  author={Bailey, Diane E and Kurland, Nancy B},
  journal={Journal of Organizational Behavior: The International Journal of Industrial, Occupational and Organizational Psychology and Behavior},
  volume={23},
  number={4},
  pages={383--400},
  year={2002},
  publisher={Wiley Online Library}
}

@article{felstead2017assessing,
  title={Assessing the growth of remote working and its consequences for effort, well-being and work-life balance},
  author={Felstead, Alan and Henseke, Golo},
  journal={New Technology, Work and Employment},
  volume={32},
  number={3},
  pages={195--212},
  year={2017},
  publisher={Wiley Online Library}
}

@article{beckel2022telework,
  title={Telework and worker health and well-being: A review and recommendations for research and practice},
  author={Beckel, Julia LO and Fisher, Gwenith G},
  journal={International journal of environmental research and public health},
  volume={19},
  number={7},
  pages={3879},
  year={2022},
  publisher={MDPI}
}

@article{mazmanian2013autonomy,
  title={The autonomy paradox: The implications of mobile email devices for knowledge professionals},
  author={Mazmanian, Melissa and Orlikowski, Wanda J and Yates, JoAnne},
  journal={Organization science},
  volume={24},
  number={5},
  pages={1337--1357},
  year={2013},
  publisher={Informs}
}

@inproceedings{saha2023time,
author = {Saha, Koustuv and Iqbal, Shamsi T},
title = {Focus Time: Effectiveness of Computer Assisted Protected Time for Wellbeing and Work Engagement of Information Workers},
year = {2023},
isbn = {9798400708077},
publisher = {Association for Computing Machinery},
address = {New York, NY, USA},
url = {https://doi.org/10.1145/3596671.3598571},
doi = {10.1145/3596671.3598571},
booktitle = {Proceedings of the 2nd Annual Meeting of the Symposium on Human-Computer Interaction for Work},
articleno = {7},
numpages = {13},
location = {Oldenburg, Germany},
series = {CHIWORK '23}
}

@book{nippert2008home,
  title={Home and work: Negotiating boundaries through everyday life},
  author={Nippert-Eng, Christena E},
  year={2008},
  publisher={University of Chicago Press}
}

@article{thomson2013information,
  title={“When i've packed it in and they send me something…”: Information boundaries in professional home offices},
  author={Thomson, Leslie},
  journal={Proceedings of the American Society for Information Science and Technology},
  volume={50},
  number={1},
  pages={1--5},
  year={2013},
  publisher={Wiley Online Library}
}

@article{steup2022attune,
author = {Steup, Rosemary and White, Paige and Dombrowski, Lynn and Su, Norman Makoto},
title = {"A Reasonable Life": Rhythmic Attunement and Sustainable Work at the Intersection of Farming and Knowledge Work},
year = {2022},
issue_date = {November 2022},
publisher = {Association for Computing Machinery},
address = {New York, NY, USA},
volume = {6},
number = {CSCW2},
url = {https://doi.org/10.1145/3555577},
doi = {10.1145/3555577},
journal = {Proc. ACM Hum.-Comput. Interact.},
month = nov,
articleno = {476},
numpages = {22}
}

@article{jarrahi2022digital,
  title={Digital assemblages, information infrastructures, and mobile knowledge work},
  author={Jarrahi, Mohammad Hossein and Sawyer, Steve and Erickson, Ingrid},
  journal={Journal of Information Technology},
  volume={37},
  number={3},
  pages={230--249},
  year={2022},
  publisher={SAGE Publications Sage UK: London, England}
}

@inproceedings{mehrvarz2025visibility,
author = {Mehrvarz, Mahan and Murray-Rust, Dave and Verma, Himanshu and Wagner, Ben},
title = {Framing the (in)visible: Insights into Visibility Practices of Remote Knowledge Workers},
year = {2025},
isbn = {9798400713842},
publisher = {Association for Computing Machinery},
address = {New York, NY, USA},
url = {https://doi.org/10.1145/3729176.3729178},
doi = {10.1145/3729176.3729178},
booktitle = {Proceedings of the 4th Annual Symposium on Human-Computer Interaction for Work},
articleno = {1},
numpages = {14},
location = {
},
series = {CHIWORK '25}
}

@article{olson2000distance,
  title={Distance matters},
  author={Olson, Gary M and Olson, Judith S},
  journal={Human--computer interaction},
  volume={15},
  number={2-3},
  pages={139--178},
  year={2000},
  publisher={Taylor \& Francis}
}

@article{walton1973quality,
  title={Quality of working life: what is it?},
  author={Walton, Richard E},
  journal={Sloan Management Review (pre-1986)},
  volume={15},
  number={1},
  pages={11},
  year={1973},
  publisher={Massachusetts Institute of Technology, Cambridge, MA}
}

@incollection{marcus2023covid,
  title={COVID-19 and the shift to remote work},
  author={Marcus, J Scott},
  booktitle={Beyond the Pandemic? Exploring the Impact of COVID-19 on Telecommunications and the Internet},
  pages={71--102},
  year={2023},
  publisher={Emerald Publishing Limited}
}

@inproceedings{cecchinato2015micro,
author = {Cecchinato, Marta E. and Cox, Anna L. and Bird, Jon},
title = {Working 9-5? Professional Differences in Email and Boundary Management Practices},
year = {2015},
isbn = {9781450331456},
publisher = {Association for Computing Machinery},
address = {New York, NY, USA},
url = {https://doi.org/10.1145/2702123.2702537},
doi = {10.1145/2702123.2702537},
booktitle = {Proceedings of the 33rd Annual ACM Conference on Human Factors in Computing Systems},
pages = {3989–3998},
numpages = {10},
location = {Seoul, Republic of Korea},
series = {CHI '15}
}

@inproceedings{pei2019we,
author = {Pei, Lucy and Nardi, Bonnie},
title = {We Did It Right, But It Was Still Wrong: Toward Assets-Based Design},
year = {2019},
isbn = {9781450359719},
publisher = {Association for Computing Machinery},
address = {New York, NY, USA},
url = {https://doi.org/10.1145/3290607.3310434},
doi = {10.1145/3290607.3310434},
booktitle = {Extended Abstracts of the 2019 CHI Conference on Human Factors in Computing Systems},
pages = {1–11},
numpages = {11},
location = {Glasgow, Scotland Uk},
series = {CHI EA '19}
}

@article{meda2016future,
  title={The future of work: the meaning and value of work in Europe},
  author={M{\'e}da, Dominique},
  year={2016}
}

@article{morse1955function,
  title={The function and meaning of work and the job},
  author={Morse, Nancy C and Weiss, Robert S},
  journal={American sociological review},
  volume={20},
  number={2},
  pages={191--198},
  year={1955},
  publisher={JSTOR}
}

@article{williams2016beyond,
  title={Beyond work-life “integration”},
  author={Williams, Joan C and Berdahl, Jennifer L and Vandello, Joseph A},
  journal={Annual review of psychology},
  volume={67},
  number={1},
  pages={515--539},
  year={2016},
  publisher={Annual Reviews}
}

@book{beckman2020dreams,
  title={Dreams of the overworked: Living, working, and parenting in the digital age},
  author={Beckman, Christine M and Mazmanian, Melissa},
  year={2020},
  publisher={Stanford University Press},
  address={Stanford, CA}
}

@inproceedings{erickson2016infra,
author = {Erickson, Ingrid and Jarrahi, Mohammad Hossein},
title = {Infrastructuring and the Challenge of Dynamic Seams in Mobile Knowledge Work},
year = {2016},
isbn = {9781450335928},
publisher = {Association for Computing Machinery},
address = {New York, NY, USA},
url = {https://doi.org/10.1145/2818048.2820015},
doi = {10.1145/2818048.2820015},
booktitle = {Proceedings of the 19th ACM Conference on Computer-Supported Cooperative Work \& Social Computing},
pages = {1323–1336},
numpages = {14},
location = {San Francisco, California, USA},
series = {CSCW '16}
}

@article{vega2023platform,
author = {Alvarez de la Vega, Juan Carlos and Cecchinato, Marta E. and Rooksby, John and Newbold, Joseph},
title = {Understanding Platform Mediated Work-Life: A Diary Study with Gig Economy Freelancers},
year = {2023},
issue_date = {April 2023},
publisher = {Association for Computing Machinery},
address = {New York, NY, USA},
volume = {7},
number = {CSCW1},
url = {https://doi.org/10.1145/3579539},
doi = {10.1145/3579539},
journal = {Proc. ACM Hum.-Comput. Interact.},
month = apr,
articleno = {106},
numpages = {32}
}

@article{erickson2023optimizing,
  title={Optimizing temporal capital: How big tech imagines time as auditable},
  author={Erickson, Ingrid and Wajcman, Judy},
  journal={American Behavioral Scientist},
  volume={67},
  number={14},
  pages={1755--1770},
  year={2023},
  publisher={SAGE Publications Sage CA: Los Angeles, CA}
}

@inproceedings{mazmanian2015circumscribed,
author = {Mazmanian, Melissa and Erickson, Ingrid and Harmon, Ellie},
title = {Circumscribed Time and Porous Time: Logics as a Way of Studying Temporality},
year = {2015},
isbn = {9781450329224},
publisher = {Association for Computing Machinery},
address = {New York, NY, USA},
url = {https://doi-org.libproxy1.nus.edu.sg/10.1145/2675133.2675231},
doi = {10.1145/2675133.2675231},
booktitle = {Proceedings of the 18th ACM Conference on Computer Supported Cooperative Work \& Social Computing},
pages = {1453–1464},
numpages = {12},
location = {Vancouver, BC, Canada},
series = {CSCW '15}
}

@article{breideband2022rhythm,
author = {Breideband, Thomas and Talkad Sukumar, Poorna and Mark, Gloria and Caruso, Megan and D'Mello, Sidney and Striegel, Aaron D.},
title = {Home-Life and Work Rhythm Diversity in Distributed Teamwork: A Study with Information Workers during the COVID-19 Pandemic},
year = {2022},
issue_date = {April 2022},
publisher = {Association for Computing Machinery},
address = {New York, NY, USA},
volume = {6},
number = {CSCW1},
url = {https://doi-org.libproxy1.nus.edu.sg/10.1145/3512942},
doi = {10.1145/3512942},
journal = {Proc. ACM Hum.-Comput. Interact.},
month = apr,
articleno = {95},
numpages = {23}
}

@article{jarrahi2021flexible,
author = {Jarrahi, Mohammad Hossein and Newlands, Gemma and Butler, Brian and Savage, Saiph and Lutz, Christoph and Dunn, Michael and Sawyer, Steve},
title = {Flexible work and personal digital infrastructures},
year = {2021},
issue_date = {July 2021},
publisher = {Association for Computing Machinery},
address = {New York, NY, USA},
volume = {64},
number = {7},
issn = {0001-0782},
url = {https://doi-org.libproxy1.nus.edu.sg/10.1145/3419405},
doi = {10.1145/3419405},
journal = {Commun. ACM},
month = jun,
pages = {72–79},
numpages = {8}
}

@article{mckie2002shadow,
  title={Shadow times: The temporal and spatial frameworks and experiences of caring and working},
  author={McKie, Linda and Gregory, Susan and Bowlby, Sophia},
  journal={Sociology},
  volume={36},
  number={4},
  pages={897--924},
  year={2002},
  publisher={Sage Publications Sage CA: Thousand Oaks, CA}
}

@inproceedings{ciolfi2020making,
  title={Making home work places},
  author={Ciolfi, Luigina and Gray, Breda and Pinatti De Carvalho, Aparecido Fabiano},
  booktitle={Proceedings of 18th European Conference on Computer-Supported Cooperative Work},
  volume={4},
  number={1},
  year={2020},
  organization={European Society for Socially Embedded Technologies}
}

@article{bodker2016rethinking,
  title={Rethinking technology on the boundaries of life and work},
  author={B{\o}dker, Susanne},
  journal={Personal and Ubiquitous Computing},
  volume={20},
  number={4},
  pages={533--544},
  year={2016},
  publisher={Springer}
}

@article{messenger2016three,
  title={Three generations of Telework: New ICT s and the (R) evolution from Home Office to Virtual Office},
  author={Messenger, Jon C and Gschwind, Lutz},
  journal={New Technology, Work and Employment},
  volume={31},
  number={3},
  pages={195--208},
  year={2016},
  publisher={Wiley Online Library}
}

@incollection{kossek2011flexible,
  author    = {Kossek, Ellen Ernst and Michel, Jesse S.},
  title     = {Flexible work schedules},
  booktitle = {APA Handbook of Industrial and Organizational Psychology, Vol. 1: Building and Developing the Organization},
  editor    = {Zedeck, Sheldon},
  publisher = {American Psychological Association},
  address   = {Washington, DC},
  year      = {2011},
  pages     = {535--572},
  doi       = {10.1037/12169-017}
}

@article{williams2023boundary,
author = {Williams, Alex C. and Iqbal, Shamsi and Kiseleva, Julia and White, Ryen W.},
title = {Managing Tasks across the Work–Life Boundary: Opportunities, Challenges, and Directions},
year = {2023},
issue_date = {June 2023},
publisher = {Association for Computing Machinery},
address = {New York, NY, USA},
volume = {30},
number = {3},
issn = {1073-0516},
url = {https://doi.org/10.1145/3582429},
doi = {10.1145/3582429},
journal = {ACM Trans. Comput.-Hum. Interact.},
month = jun,
articleno = {48},
numpages = {31}
}

@inproceedings{lee2006human,
author = {Lee, Charlotte P. and Dourish, Paul and Mark, Gloria},
title = {The human infrastructure of cyberinfrastructure},
year = {2006},
isbn = {1595932496},
publisher = {Association for Computing Machinery},
address = {New York, NY, USA},
url = {https://doi-org.libproxy1.nus.edu.sg/10.1145/1180875.1180950},
doi = {10.1145/1180875.1180950},
booktitle = {Proceedings of the 2006 20th Anniversary Conference on Computer Supported Cooperative Work},
pages = {483–492},
numpages = {10},
location = {Banff, Alberta, Canada},
series = {CSCW '06}
}

@inproceedings{avrahami2020celebrating,
author = {Avrahami, Daniel and Williams, Kristin and Lee, Matthew L. and Tokunaga, Nami and Tjahjadi, Yulius and Marlow, Jennifer},
title = {Celebrating Everyday Success: Improving Engagement and Motivation using a System for Recording Daily Highlights},
year = {2020},
isbn = {9781450367080},
publisher = {Association for Computing Machinery},
address = {New York, NY, USA},
url = {https://doi.org/10.1145/3313831.3376369},
doi = {10.1145/3313831.3376369},
booktitle = {Proceedings of the 2020 CHI Conference on Human Factors in Computing Systems},
pages = {1–13},
numpages = {13},
location = {Honolulu, HI, USA},
series = {CHI '20}
}

@inproceedings{schorch2016designing,
author = {Schorch, Mar\'{e}n and Wan, Lin and Randall, David William and Wulf, Volker},
title = {Designing for Those who are Overlooked: Insider Perspectives on Care Practices and Cooperative Work of Elderly Informal Caregivers},
year = {2016},
isbn = {9781450335928},
publisher = {Association for Computing Machinery},
address = {New York, NY, USA},
url = {https://doi.org/10.1145/2818048.2819999},
doi = {10.1145/2818048.2819999},
booktitle = {Proceedings of the 19th ACM Conference on Computer-Supported Cooperative Work \& Social Computing},
pages = {787–799},
numpages = {13},
location = {San Francisco, California, USA},
series = {CSCW '16}
}

@article{chow2023feeling,
author = {Chow, Kevin and Fritz, Thomas and Holsti, Liisa and Barbic, Skye and McGrenere, Joanna},
title = {Feeling Stressed and Unproductive? A Field Evaluation of a Therapy-Inspired Digital Intervention for Knowledge Workers},
year = {2023},
issue_date = {February 2024},
publisher = {Association for Computing Machinery},
address = {New York, NY, USA},
volume = {31},
number = {1},
issn = {1073-0516},
url = {https://doi.org/10.1145/3609330},
doi = {10.1145/3609330},
journal = {ACM Trans. Comput.-Hum. Interact.},
month = nov,
articleno = {12},
numpages = {33}
}

@inproceedings{reddy2002pulse,
author = {Reddy, Madhu and Dourish, Paul},
title = {A finger on the pulse: temporal rhythms and information seeking in medical work},
year = {2002},
isbn = {1581135602},
publisher = {Association for Computing Machinery},
address = {New York, NY, USA},
url = {https://doi-org.libproxy1.nus.edu.sg/10.1145/587078.587126},
doi = {10.1145/587078.587126},
booktitle = {Proceedings of the 2002 ACM Conference on Computer Supported Cooperative Work},
pages = {344–353},
numpages = {10},
location = {New Orleans, Louisiana, USA},
series = {CSCW '02}
}

@inproceedings{nilsson2005negotiated,
author = {Nilsson, Magnus and Hertzum, Morten},
title = {Negotiated rhythms of mobile work: time, place, and work schedules},
year = {2005},
isbn = {1595932232},
publisher = {Association for Computing Machinery},
address = {New York, NY, USA},
url = {https://doi-org.libproxy1.nus.edu.sg/10.1145/1099203.1099233},
doi = {10.1145/1099203.1099233},
booktitle = {Proceedings of the 2005 ACM International Conference on Supporting Group Work},
pages = {148–157},
numpages = {10},
location = {Sanibel Island, Florida, USA},
series = {GROUP '05}
}

@article{sun2024mixed,
author = {Sun, Lu and Mok, Lillio and Sen, Shilad and Sarrafzadeh, Bahareh},
title = {Rhythm of Work: Mixed-methods Characterization of Information Workers Scheduling Preferences and Practices},
year = {2024},
issue_date = {November 2024},
publisher = {Association for Computing Machinery},
address = {New York, NY, USA},
volume = {8},
number = {CSCW2},
url = {https://doi.org/10.1145/3686895},
doi = {10.1145/3686895},
journal = {Proc. ACM Hum.-Comput. Interact.},
month = nov,
articleno = {356},
numpages = {38}
}

@inproceedings{mazmanian2005crackberries,
  title={Crackberries: The social implications of ubiquitous wireless e-mail devices},
  author={Mazmanian, Melissa A and Orlikowski, Wanda J and Yates, JoAnne},
  booktitle={Designing Ubiquitous Information Environments: Socio-Technical Issues and Challenges: IFIP TC8 WG 8.2 International Working Conference, August 1--3, 2005, Cleveland, Ohio, USA},
  pages={337--343},
  year={2005},
  organization={Springer}
}

@article{voydanoff2004effects,
  title={The effects of work demands and resources on work-to-family conflict and facilitation},
  author={Voydanoff, Patricia},
  journal={Journal of Marriage and family},
  volume={66},
  number={2},
  pages={398--412},
  year={2004},
  publisher={Wiley Online Library}
}

@article{greenhaus1985sources,
  title={Sources of conflict between work and family roles},
  author={Greenhaus, Jeffrey H and Beutell, Nicholas J},
  journal={Academy of management review},
  volume={10},
  number={1},
  pages={76--88},
  year={1985},
  publisher={Academy of management Briarcliff Manor, NY 10510}
}

@article{greenhaus2006work,
  title={When work and family are allies: A theory of work-family enrichment},
  author={Greenhaus, Jeffrey H and Powell, Gary N},
  journal={Academy of management review},
  volume={31},
  number={1},
  pages={72--92},
  year={2006},
  publisher={Academy of Management Briarcliff Manor, NY 10510}
}

@article{alvarez2023understanding,
  title={Understanding platform mediated work-life: a diary study with gig economy freelancers},
  author={Alvarez de la Vega, Juan Carlos and Cecchinato, Marta E and Rooksby, John and Newbold, Joseph},
  journal={Proceedings of the ACM on Human-Computer Interaction},
  volume={7},
  number={CSCW1},
  pages={1--32},
  year={2023},
  publisher={ACM New York, NY, USA}
}

@article{grant2013exploration,
  title={An exploration of the psychological factors affecting remote e-worker's job effectiveness, well-being and work-life balance},
  author={Grant, Christine A and Wallace, Louise M and Spurgeon, Peter C},
  journal={Employee relations},
  volume={35},
  number={5},
  pages={527--546},
  year={2013},
  publisher={Emerald Group Publishing Limited}
}

@article{bernhardt2023work,
  title={Work from home and parenting: Examining the role of work-family conflict and gender during the COVID-19 pandemic},
  author={Bernhardt, Janine and Recksiedler, Claudia and Linberg, Anja},
  journal={Journal of social issues},
  volume={79},
  number={3},
  pages={935--970},
  year={2023},
  publisher={Wiley Online Library}
}

@inproceedings{epstein2016taking,
author = {Epstein, Daniel A. and Avrahami, Daniel and Biehl, Jacob T.},
title = {Taking 5: Work-Breaks, Productivity, and Opportunities for Personal Informatics for Knowledge Workers},
year = {2016},
isbn = {9781450333627},
publisher = {Association for Computing Machinery},
address = {New York, NY, USA},
url = {https://doi.org/10.1145/2858036.2858066},
doi = {10.1145/2858036.2858066},
booktitle = {Proceedings of the 2016 CHI Conference on Human Factors in Computing Systems},
pages = {673–684},
numpages = {12},
location = {San Jose, California, USA},
series = {CHI '16}
}

@inproceedings{cambo2017break,
author = {Cambo, Scott A. and Avrahami, Daniel and Lee, Matthew L.},
title = {BreakSense: Combining Physiological and Location Sensing to Promote Mobility during Work-Breaks},
year = {2017},
isbn = {9781450346559},
publisher = {Association for Computing Machinery},
address = {New York, NY, USA},
url = {https://doi-org.libproxy1.nus.edu.sg/10.1145/3025453.3026021},
doi = {10.1145/3025453.3026021},
booktitle = {Proceedings of the 2017 CHI Conference on Human Factors in Computing Systems},
pages = {3595–3607},
numpages = {13},
location = {Denver, Colorado, USA},
series = {CHI '17}
}

@inproceedings{howe2022microbreak,
author = {Howe, Esther and Suh, Jina and Bin Morshed, Mehrab and McDuff, Daniel and Rowan, Kael and Hernandez, Javier and Abdin, Marah Ihab and Ramos, Gonzalo and Tran, Tracy and Czerwinski, Mary P},
title = {Design of Digital Workplace Stress-Reduction Intervention Systems: Effects of Intervention Type and Timing},
year = {2022},
isbn = {9781450391573},
publisher = {Association for Computing Machinery},
address = {New York, NY, USA},
url = {https://doi.org/10.1145/3491102.3502027},
doi = {10.1145/3491102.3502027},
booktitle = {Proceedings of the 2022 CHI Conference on Human Factors in Computing Systems},
articleno = {327},
numpages = {16},
location = {New Orleans, LA, USA},
series = {CHI '22}
}

@inproceedings{cho2024feminist,
author = {Cho, Janghee and Voida, Stephen},
title = {Toward More Inclusive and Accessible Futures of Remote Work Using a Feminist Geographical Lens},
year = {2024},
isbn = {9798400710179},
publisher = {Association for Computing Machinery},
address = {New York, NY, USA},
url = {https://doi.org/10.1145/3663384.3663392},
doi = {10.1145/3663384.3663392},
booktitle = {Proceedings of the 3rd Annual Meeting of the Symposium on Human-Computer Interaction for Work},
articleno = {4},
numpages = {10},
location = {Newcastle upon Tyne, United Kingdom},
series = {CHIWORK '24}
}

@article{poelmans2008achieving,
  title={Achieving work--life balance: Current theoretical and practice issues},
  author={Poelmans, Steven AY and Kalliath, Thomas and Brough, Paula},
  journal={Journal of Management \& Organization},
  volume={14},
  number={3},
  pages={227--238},
  year={2008},
  publisher={Cambridge University Press}
}

@article{kalliath2008work,
  title={Work--life balance: A review of the meaning of the balance construct},
  author={Kalliath, Thomas and Brough, Paula},
  journal={Journal of management \& organization},
  volume={14},
  number={3},
  pages={323--327},
  year={2008},
  publisher={Cambridge University Press}
}

@article{de2025living,
  title={Living to work (from home): Overwork, remote work, and gendered dual devotion to work and family},
  author={de Laat, Kim},
  journal={Work and Occupations},
  volume={52},
  number={1},
  pages={130--165},
  year={2025},
  publisher={SAGE Publications Sage CA: Los Angeles, CA}
}

@inproceedings{das2023focused,
author = {Das Swain, Vedant and Hernandez, Javier and Houck, Brian and Saha, Koustuv and Suh, Jina and Chaudhry, Ahad and Cho, Tenny and Guo, Wendy and Iqbal, Shamsi and Czerwinski, Mary P},
title = {Focused Time Saves Nine: Evaluating Computer–Assisted Protected Time for Hybrid Information Work},
year = {2023},
isbn = {9781450394215},
publisher = {Association for Computing Machinery},
address = {New York, NY, USA},
url = {https://doi.org/10.1145/3544548.3581326},
doi = {10.1145/3544548.3581326},
booktitle = {Proceedings of the 2023 CHI Conference on Human Factors in Computing Systems},
articleno = {857},
numpages = {18},
location = {Hamburg, Germany},
series = {CHI '23}
}

@inproceedings{das2022two,
author = {Das Swain, Vedant and Williams, Shane and Fourney, Adam and Iqbal, Shamsi T.},
title = {Two Birds with One Phone: The Role of Mobile Use in the Daily Practices of Remote Information Work},
year = {2022},
isbn = {9781450396554},
publisher = {Association for Computing Machinery},
address = {New York, NY, USA},
url = {https://doi.org/10.1145/3533406.3533416},
doi = {10.1145/3533406.3533416},
booktitle = {Proceedings of the 1st Annual Meeting of the Symposium on Human-Computer Interaction for Work},
articleno = {2},
numpages = {8},
location = {Durham, NH, USA},
series = {CHIWORK '22}
}

@article{allen2013work,
  title={Work--family conflict and flexible work arrangements: Deconstructing flexibility},
  author={Allen, Tammy D and Johnson, Ryan C and Kiburz, Kaitlin M and Shockley, Kristen M},
  journal={Personnel psychology},
  volume={66},
  number={2},
  pages={345--376},
  year={2013},
  publisher={Wiley Online Library}
}

@article{mols2021always,
  title={Always available via WhatsApp: Mapping everyday boundary work practices and privacy negotiations},
  author={Mols, Anouk and Pridmore, Jason},
  journal={Mobile Media \& Communication},
  volume={9},
  number={3},
  pages={422--440},
  year={2021},
  publisher={SAGE Publications Sage UK: London, England}
}

@article{kossek2025reenvisioning,
  title={Reenvisioning Family-Supportive Organizations Through a Diversity, Equity, and Inclusion Perspective: A Review and Research Agenda},
  author={Kossek, Ellen Ernst and Vaziri, Hoda and Perrigino, Matthew B and Lautsch, Brenda A and Pratt, Benjamin R and King, Eden B},
  journal={Journal of Management},
  volume={51},
  number={6},
  pages={2520--2548},
  year={2025},
  publisher={SAGE Publications Sage CA: Los Angeles, CA}
}

@book{duxbury2009balancing,
  title={Balancing paid work and caregiving responsibilities: A closer look at family caregivers in Canada},
  author={Duxbury, Linda Elizabeth and Higgins, Christopher Alan and Schroeder, Bonnie},
  year={2009},
  publisher={Canadian Policy Research Networks Ottawa, ON, Canada}
}

@article{clark2000work,
  title={Work/family border theory: A new theory of work/family balance},
  author={Clark, Sue Campbell},
  journal={Human relations},
  volume={53},
  number={6},
  pages={747--770},
  year={2000},
  publisher={Sage Publications Sage CA: Thousand Oaks, CA}
}

@article{ashforth2000all,
  title={All in a day's work: Boundaries and micro role transitions},
  author={Ashforth, Blake E and Kreiner, Glen E and Fugate, Mel},
  journal={Academy of Management review},
  volume={25},
  number={3},
  pages={472--491},
  year={2000},
  publisher={Academy of Management Briarcliff Manor, NY 10510}
}

@book{zerubavel1993fine,
  title={The fine line},
  author={Zerubavel, Eviatar},
  year={1993},
  publisher={University of Chicago Press}
}

@article{kretzmann1996assets,
  title={Assets-based community development},
  author={Kretzmann, John and McKnight, John P},
  journal={Nat'l Civic Rev.},
  volume={85},
  pages={23},
  year={1996},
  publisher={HeinOnline}
}

@article{mathie2003clients,
  title={From clients to citizens: Asset-based community development as a strategy for community-driven development},
  author={Mathie, Alison and Cunningham, Gord},
  journal={Development in practice},
  volume={13},
  number={5},
  pages={474--486},
  year={2003},
  publisher={Taylor \& Francis}
}

@inproceedings{cho2019comadre,
author = {Cho, Alexander and Herrera, Roxana G. and Chaidez, Luis and Uriostegui, Adilene},
title = {The "Comadre" Project: An Asset-Based Design Approach to Connecting Low-Income Latinx Families to Out-of-School Learning Opportunities},
year = {2019},
isbn = {9781450359702},
publisher = {Association for Computing Machinery},
address = {New York, NY, USA},
url = {https://doi.org/10.1145/3290605.3300837},
doi = {10.1145/3290605.3300837},
booktitle = {Proceedings of the 2019 CHI Conference on Human Factors in Computing Systems},
pages = {1–14},
numpages = {14},
location = {Glasgow, Scotland Uk},
series = {CHI '19}
}

@inproceedings{ahumada2021call,
author = {Ahumada-Newhart, Veronica and Maya Hernandez, J. and Badillo-Urquiola, Karla},
title = {A Call for Action: Conceptualizing Assets-Based Inclusive Design as a Social Movement to Address Systemic Inequities: An Assets-Based Inclusive Design Framework},
year = {2021},
isbn = {9781450380959},
publisher = {Association for Computing Machinery},
address = {New York, NY, USA},
url = {https://doi.org/10.1145/3411763.3450368},
doi = {10.1145/3411763.3450368},
booktitle = {Extended Abstracts of the 2021 CHI Conference on Human Factors in Computing Systems},
articleno = {12},
numpages = {4},
location = {Yokohama, Japan},
series = {CHI EA '21}
}

@inproceedings{gautam2020crafting,
author = {Gautam, Aakash and Tatar, Deborah and Harrison, Steve},
title = {Crafting, Communality, and Computing: Building on Existing Strengths To Support a Vulnerable Population},
year = {2020},
isbn = {9781450367080},
publisher = {Association for Computing Machinery},
address = {New York, NY, USA},
url = {https://doi.org/10.1145/3313831.3376647},
doi = {10.1145/3313831.3376647},
booktitle = {Proceedings of the 2020 CHI Conference on Human Factors in Computing Systems},
pages = {1–14},
numpages = {14},
location = {Honolulu, HI, USA},
series = {CHI '20}
}

@inproceedings{irani2018refuge,
author = {Irani, Azalea and Nelavelli, Kriti and Hare, Kristin and Bondal, Paula and Kumar, Neha},
title = {Refuge Tech: An Assets-Based Approach to Refugee Resettlement},
year = {2018},
isbn = {9781450356213},
publisher = {Association for Computing Machinery},
address = {New York, NY, USA},
url = {https://doi.org/10.1145/3170427.3188680},
doi = {10.1145/3170427.3188680},
booktitle = {Extended Abstracts of the 2018 CHI Conference on Human Factors in Computing Systems},
pages = {1–6},
numpages = {6},
location = {Montreal QC, Canada},
series = {CHI EA '18}
}

@article{wong2021needs,
author = {Wong-Villacres, Marisol and Gautam, Aakash and Tatar, Deborah and DiSalvo, Betsy},
title = {Reflections on Assets-Based Design: A Journey Towards A Collective of Assets-Based Thinkers},
year = {2021},
issue_date = {October 2021},
publisher = {Association for Computing Machinery},
address = {New York, NY, USA},
volume = {5},
number = {CSCW2},
url = {https://doi.org/10.1145/3479545},
doi = {10.1145/3479545},
journal = {Proc. ACM Hum.-Comput. Interact.},
month = oct,
articleno = {401},
numpages = {32}
}

@inproceedings{petterson2024networks,
author = {Petterson, Adrian and Jaimes Rodriguez, Isabella and Doggett, Olivia and Chandra, Priyank},
title = {Networks of care in digital domestic labour economies},
year = {2024},
isbn = {9798400703300},
publisher = {Association for Computing Machinery},
address = {New York, NY, USA},
url = {https://doi.org/10.1145/3613904.3642200},
doi = {10.1145/3613904.3642200},
booktitle = {Proceedings of the 2024 CHI Conference on Human Factors in Computing Systems},
articleno = {530},
numpages = {16},
location = {Honolulu, HI, USA},
series = {CHI '24}
}

@inproceedings{guillou2020your,
author = {Guillou, Hayley and Chow, Kevin and Fritz, Thomas and McGrenere, Joanna},
title = {Is Your Time Well Spent? Reflecting on Knowledge Work More Holistically},
year = {2020},
isbn = {9781450367080},
publisher = {Association for Computing Machinery},
address = {New York, NY, USA},
url = {https://doi.org/10.1145/3313831.3376586},
doi = {10.1145/3313831.3376586},
booktitle = {Proceedings of the 2020 CHI Conference on Human Factors in Computing Systems},
pages = {1–9},
numpages = {9},
location = {Honolulu, HI, USA},
series = {CHI '20}
}

@article{wong2014paradigm,
  title={A paradigm shift in regulating and running nursing homes in Singapore},
  author={Wong, Gabriel HZ and Pang, Weng Sun and Yap, Philip},
  journal={Journal of the American Medical Directors Association},
  volume={15},
  number={6},
  pages={440--444},
  year={2014},
  publisher={Elsevier}
}

@book{ho2018care,
  title={Care where you are: Enabling Singaporeans to age well in the community},
  author={Ho, Elaine Lynn-Ee and Huang, Shirlena},
  year={2018},
  publisher={Straits Times Press Pte Limited}
}

@article{yeoh1999migrant,
  title={Migrant female domestic workers: debating the economic, social and political impacts in Singapore},
  author={Yeoh, Brenda SA and Huang, Shirlena and Gonzalez III, Joaquin},
  journal={International Migration Review},
  volume={33},
  number={1},
  pages={114--136},
  year={1999},
  publisher={Sage Publications Sage CA: Los Angeles, CA}
}

@online{ntuc2025familyfriendly,
  author       = {{National Trades Union Congress (NTUC)}},
  title        = {Family-Friendly Workplaces in Singapore and What to Look For},
  year         = {2025},
  month        = jun # "~5",
  howpublished = {\url{https://www.ntuc.org.sg/sube/news/family-friendly-workplaces-in-singapore-and-what-to-look-for/}},
  note         = {Accessed: 2025-09-09}
}

@article{tan2006influence,
  title={The influence of value orientations and demographics on quality-of-life perceptions: Evidence from a national survey of Singaporeans},
  author={Tan, Soo Jiuan and Tambyah, Siok Kuan and Kau, Ah Keng},
  journal={Social Indicators Research},
  volume={78},
  number={1},
  pages={33--59},
  year={2006},
  publisher={Springer}
}

@article{kossek2021future,
  title={The future of flexibility at work},
  author={Kossek, Ellen Ernst and Gettings, Patricia and Misra, Kaumudi},
  journal={Harvard Business Review},
  volume={28},
  year={2021}
}

@article{teo2018whose,
  title={Whose Family Matters? Work--Care--Migration Regimes and Class Inequalities in Singapore},
  author={Teo, Youyenn},
  journal={Critical Sociology},
  volume={44},
  number={7-8},
  pages={1133--1146},
  year={2018},
  publisher={SAGE Publications Sage UK: London, England}
}

@article{kossek2023making,
  title={Making flexibility more i-deal: Advancing work-life equality collectively},
  author={Kossek, Ellen Ernst and Kelliher, Clare},
  journal={Group \& Organization Management},
  volume={48},
  number={1},
  pages={317--349},
  year={2023},
  publisher={Sage Publications Sage CA: Los Angeles, CA}
}

@book{costanza2020design,
  title={Design justice: Community-led practices to build the worlds we need},
  author={Costanza-Chock, Sasha},
  year={2020},
  publisher={The MIT Press}
}

@article{wong2004spaces,
  title={Spaces of silence: single parenthood and the ‘normal family’in Singapore},
  author={Wong, Theresa and Yeoh, Brenda SA and Graham, Elspeth F and Teo, Peggy},
  journal={Population, Space and Place},
  volume={10},
  number={1},
  pages={43--58},
  year={2004},
  publisher={Wiley Online Library}
}

@article{pschetz2015time,
author = {Pschetz, Larissa},
title = {Isn't it time to change the way we think about time?},
year = {2015},
issue_date = {September-October 2015},
publisher = {Association for Computing Machinery},
address = {New York, NY, USA},
volume = {22},
number = {5},
issn = {1072-5520},
url = {https://doi.org/10.1145/2809502},
doi = {10.1145/2809502},
journal = {Interactions},
month = aug,
pages = {58–61},
numpages = {4}
}

@inproceedings{snyder2019bipolar,
author = {Snyder, Jaime and Murnane, Elizabeth and Lustig, Caitie and Voida, Stephen},
title = {Visually Encoding the Lived Experience of Bipolar Disorder},
year = {2019},
isbn = {9781450359702},
publisher = {Association for Computing Machinery},
address = {New York, NY, USA},
url = {https://doi.org/10.1145/3290605.3300363},
doi = {10.1145/3290605.3300363},
booktitle = {Proceedings of the 2019 CHI Conference on Human Factors in Computing Systems},
pages = {1–14},
numpages = {14},
location = {Glasgow, Scotland Uk},
series = {CHI '19}
}

@inproceedings{alfaras2020somadata, author = {Alfaras, Miquel and Tsaknaki, Vasiliki and Sanches, Pedro and Windlin, Charles and Umair, Muhammad and Sas, Corina and H\"{o}\"{o}k, Kristina}, title = {From Biodata to Somadata}, year = {2020}, isbn = {9781450367080}, publisher = {Association for Computing Machinery}, address = {New York, NY, USA}, url = {https://doi.org/10.1145/3313831.3376684}, doi = {10.1145/3313831.3376684}, booktitle = {Proceedings of the 2020 CHI Conference on Human Factors in Computing Systems}, pages = {1–14}, numpages = {14}, location = {Honolulu, HI, USA}, series = {CHI '20} }

@book{lupi2016dear,
  title={Dear data},
  author={Lupi, Giorgia and Posavec, Stefanie},
  year={2016},
  publisher={Chronicle books}
}

@inproceedings{brock2024sonificaiton,
author = {Wirfs-Brock, Jordan and Perera, Jamie and Geere, Duncan},
title = {Open Sonifications: A Manifesto for Many Ecologies of Data and Sound},
year = {2024},
isbn = {9798400705830},
publisher = {Association for Computing Machinery},
address = {New York, NY, USA},
url = {https://doi.org/10.1145/3643834.3660757},
doi = {10.1145/3643834.3660757},
booktitle = {Proceedings of the 2024 ACM Designing Interactive Systems Conference},
pages = {2660–2674},
numpages = {15},
location = {Copenhagen, Denmark},
series = {DIS '24}
}

@inproceedings{thudt2018reflection,
author = {Thudt, Alice and Hinrichs, Uta and Huron, Samuel and Carpendale, Sheelagh},
title = {Self-Reflection and Personal Physicalization Construction},
year = {2018},
isbn = {9781450356206},
publisher = {Association for Computing Machinery},
address = {New York, NY, USA},
url = {https://doi.org/10.1145/3173574.3173728},
doi = {10.1145/3173574.3173728},
booktitle = {Proceedings of the 2018 CHI Conference on Human Factors in Computing Systems},
pages = {1–13},
numpages = {13},
location = {Montreal QC, Canada},
series = {CHI '18}
}

@inproceedings{gutierrez2017tango,
author = {Gutierrez, Francisco J. and Ochoa, Sergio F.},
title = {It Takes at Least Two to Tango: Understanding the Cooperative Nature of Elderly Caregiving in Latin America},
year = {2017},
isbn = {9781450343350},
publisher = {Association for Computing Machinery},
address = {New York, NY, USA},
url = {https://doi.org/10.1145/2998181.2998314},
doi = {10.1145/2998181.2998314},
booktitle = {Proceedings of the 2017 ACM Conference on Computer Supported Cooperative Work and Social Computing},
pages = {1618–1630},
numpages = {13},
location = {Portland, Oregon, USA},
series = {CSCW '17}
}

@inproceedings{cho2024reinforcing,
author = {Cho, Janghee and Choi, Dasom and Yu, Junnan and Voida, Stephen},
title = {Reinforcing and Reclaiming The Home: Co-speculating Future Technologies to Support Remote and Hybrid Work},
year = {2024},
isbn = {9798400703300},
publisher = {Association for Computing Machinery},
address = {New York, NY, USA},
url = {https://doi.org/10.1145/3613904.3642381},
doi = {10.1145/3613904.3642381},
booktitle = {Proceedings of the 2024 CHI Conference on Human Factors in Computing Systems},
articleno = {1026},
numpages = {28},
location = {Honolulu, HI, USA},
series = {CHI '24}
}

@misc{snef2024edb, 
  author       = "Singapore Economic Development Board ",
  title        = "Flexi-work request guidelines not meant to prescribe blanket outcomes for employers or influence hiring of workforce: SNEF",
  howpublished = "https://www.edb.gov.sg/en/business-insights/insights/flexi-work-request-guidelines-not-meant-to-prescribe-blanket-outcomes-for-employers-or-influence-hiring-of-workforce-snef.html",
  month        = "04",
  year         = "2024",
  note         = "Accessed: 2025-11-27",

}

@article{chung2020flexible,
  title={Flexible working, work--life balance, and gender equality: Introduction},
  author={Chung, Heejung and Van der Lippe, Tanja},
  journal={Social indicators research},
  volume={151},
  number={2},
  pages={365--381},
  year={2020},
  publisher={Springer}
}

@inproceedings{strahl2022making,
author = {St\r{a}hl, Anna and Balaam, Madeline and Comber, Rob and Sanches, Pedro and H\"{o}\"{o}k, Kristina},
title = {Making New Worlds – Transformative Becomings with Soma Design},
year = {2022},
isbn = {9781450391573},
publisher = {Association for Computing Machinery},
address = {New York, NY, USA},
url = {https://doi.org/10.1145/3491102.3502018},
doi = {10.1145/3491102.3502018},
booktitle = {Proceedings of the 2022 CHI Conference on Human Factors in Computing Systems},
articleno = {176},
numpages = {17},
location = {New Orleans, LA, USA},
series = {CHI '22}
}

\appendix

\section{Case Studies: Extended Accounts of Everyday Negotiations} \label{case}
Our work traced the everyday rhythms of participants as they navigated both work and care responsibilities. We drew on design sensibilities to elicit thick descriptions of their everyday lives. Through our theoretical lens (i.e., treating boundary work as assets for sustaining rhythms), we presented these experiences systematically in Section~\ref{findings}. While this structured account benefits the HCI community by clarifying how empirical insights can inform technology design, it may also simplify some of the richness of participants’ stories. To complement this, we provide case studies in this appendix that, while grounded in particular participants, are representative of themes shared across our study.

\medskip

\textbf{Case study 1}: P20 is a mom of two and a full-time graphic designer at a local magazine. Her position is fully remote, so she works from home while managing her children’s care. Her husband works long hours–he leaves by 5am for work and returns at 10pm–so it falls to her to manage the children’s care at home during the day while juggling her work responsibilities. For P20, care and work are interleaved on a typical day. She shares, \inlinequote{(I) wake up like 6.30…prepare them for school…do a little bit of housework…go to the market… 9.30, I will start work and around 12 o'clock, I'll prepare some meals before they come back home. Then I'll fetch them around 1.30… sit with them for their lunch… then do a little bit homework, maybe around 2.30 until 3.30… So after they are done with their homework, they get to do other things, then I will start to sit (at) my desk around 3 o'clock…}{P20}. Like many of our participants, P20 uses care as an organizing principle to navigate the temporal boundaries between care and work, slotting her work in between care moments.

Beyond managing time, her inner orientation towards care guides her decision making to prioritize care or work tasks at a given moment: \inlinequote{How I manage? I think I will prioritize the family, the kids first. So I'll just leave the work because as long as there's no urgency, I can always start working later or at night or I can push it for tomorrow. As long as there's no deadline, I think it's fine.}{P20}. In this way, she actively mobilizes flexibility in her time and workload as an asset, using deadlines and urgency as cues to adjust when and how she works. 

Even so, the integration of care and work boundaries creates a kind of messiness, especially during peak periods at work, creating a situation where, “the task will be haywire,” and it feels stressful and tiring. She said: \inlinequote{There is no strategy. I just try to work as much as I can during that day}{P20}. While she describes her approach as lacking strategy, her rhythm of working is in fact patterned by her values, with care foregrounded as the basis for how she gets things done and how her daily routines take shape.

During these periods, she feels that she has to sacrifice care. Yet even these moments reveal how strongly her values around care shape her choices, as she manages flexible work by adapting through substitutions, such as buying food or allowing screen time, that may not be ideal in her view but enable her to sustain both care and work:, as she puts it, \inlinequote{so I will feel that I have to sacrifice, I can't cook for them today… So I have to go out to buy food. So I have to do some replacement. I have to keep working and then they start to nag or fight. I have to scold them… or try to let them have more screen time because I can't really take care of them. So they have their own screen time. They will be keep quiet and I'll just do my work. So there'll be times when I really cannot handle their homework. So yeah, I have to sacrifice some things.}{P20}.

Despite the messiness of balancing care while working from home, P20 shares, \inlinequote{I still prefer work from home. Like how I'm doing now. Although it's very messy, but I get to spend more time with my kids and I can, I think I can do everything like how I'm doing now. I'm pretty, I feel quite balanced as well.}{P20}.

Her care priorities are further reflected in her long-term rhythm-making strategies–although she feels stagnant and wishes to develop her career, she puts her wish on hold until her children are grown to maintain her current flexible conditions for caregiving, \inlinequote{...I need the flexibility that I'm having now. So if I have to move on to upgrade myself, study or change job, I think I wouldn't be able to do what I'm doing now. Because my priority is still family and my kids... when they grow up, then I can improve on my career.}{P20}.

\medskip

\textbf{Case study 2}: P22 is a freelance editor working on a permanent part-time basis with a magazine company. She is a mother to two young children aged 10 and 6. Together with her children, she lives in a 4-bedroom flat with her husband, her two older adult parents and two domestic workers.  Her work arrangement offers complete flexibility; she can work from anywhere, with no fixed hours, as long as deadlines are met. Yet in practice, this flexibility is sharply curtailed by the crowded household, entrenched family power dynamics, and her caregiving responsibilities for children and aging parents, which shape when and how she can actually work. 

Her working spaces illustrate this tension (figure~\ref{fig:p22home}). With no dedicated office, she moves between improvised corners—the children’s playroom, the master bedroom shelf, or the living room corner. The most conducive space is her husband’s study, which has a door and air-conditioning, but it is not truly hers. Access is contingent on his absence and preferences, and she abides by unspoken rules: she leaves when he returns, avoids eating at his desk, and ensures the room is restored to his order. With eight people in one household, most spaces in the home are shared and her husband is the only member of the family with his own dedicated space. While she uses his space for work during the day when he is out, she has to adhere to his conditions, as she shares, \inlinequote{because he is the main breadwinner, usually a lot of times I would have to like defer to, to what he wants because he is after all earning most of the money. And if he says, Oh, he's very tired after, you know, workday, he wants to rest. Yeah. Then I will have to, I have to leave and go somewhere else.}{P22}. Although she finds the arrangement unfair, she accepts the arrangement to preserve household harmony. In doing so, she mobilizes what flexibility she can: rearranging her work rhythms across shifting spaces, or reserving less demanding tasks for noisier environments.  as it becomes hard to focus when working in a common area, \inlinequote{I do feel that it is a little bit unfair and it is also more difficult for me because, you know, when I'm working outside, then there'll be a lot of noise or people might feel that they can access my attention. They'll be like, “Oh, what do you think of this? What do you think of that?”}{P22}. 

While working in shared areas at home, she often has to manage her own productivity as a worker while balancing her relationship with other household members as a mother and a daughter. For example, when her parents make commentaries while watching TV, she holds back on voicing her need for focus, \inlinequote{When I'm sitting there, it's not enclosed, I can actually hear them talking… And it actually bothers me. But then when you tell them, being 89 years old and Asian parents, they feel kind of hurt… So it's very tough}{P22}. Other times, when her son is playing in the same space, “I actually would just shift off and then go to the kitchen”, she describes this as "choosing her battles", \inlinequote{I don't want to like scold him all the time, you know… I'm also quite aware that as a mother, the time is shared. I have different things to do. And a lot of times they actually with the domestic worker. So if I keep scolding him for various things, then, you know, it could affect the relationship, which is not what I want.}{P22}

Attempting to work outside the home brings its own relational tensions; as the family’s “disciplinarian”, her children take advantage of her absence to watch TV, an activity her husband disapproves of. \inlinequote{So if I leave the house and then they're watching TV the whole time at home, and then (my husband) comes back, he's not going to take well to it. So I do feel that working from home is the best option where I can actually keep an eye on certain things.}{P22}. Though restrictive, this choice illustrates how she actively mobilizes her role as a mother into a tactic that sustains her flexible work within domestic constraints. Despite these constraints, she finds ways to assert small acts of autonomy, such as finding moments of quiet at her "clean and uncluttered" kitchen counter, and consciously weaving in micro-self-care moments amid caring for others, such as indulging in fragrance samples, or sitting down in the kitchen with a cup of tea or coffee. These minor practices allow her to reclaim time and space for herself. 

\begin{quote}
    \textit{"Fragrance samples: They are a minor indulgence. I was brought up on the value that money earned had to be saved for the family. I feel bad to spend on myself sometimes. I buy the samples because I want to have nice things too, I just need to adapt my wants to my budget."}(P22, excerpt from probe kit)
\end{quote}

\medskip

\textbf{Case study 3}: P1 is a single parent raising two school-age children, one of whom has special needs. She shares her parental home with her elderly mother, aunt, and two domestic workers, while also maintaining a separate residence of her own. She alternates her time between these two households. Alongside these caregiving responsibilities, she works part-time across two organizations, one where she takes on a part-time role as a community advocate and the other one as a freelance parenting coach for employees of large multinational firms.
Her care duties for a child with special needs brings some challenges with school refusal. She recounted a time when caring for her child with special needs taxed her emotionally, \inlinequote{…I have to send him by 7.30, right?...that in itself was an intense process because he has school refusal. So it was, I really do think I have PTSD from sending him to school because every morning he is having a meltdown and then I'm having to hold space for him…And many times I have to drop him at the gate, he's crying.}{P1}.

After school drop-offs, P1 switches her focus to her work and other domains of her life. Due to her particular circumstances–managing two homes and co-parenting with her ex-husband–she is able to segment the different days of her week into distinct rhythms for deep focus work, caregiving and herself. As she shares, \inlinequote{Monday is for me to clear my work… I have a long stretch of time where I can do more, you know, complex work, work that takes more time…a bit more thinking… So Tuesday is…my daughter's day. And then Wednesday is (my son’s) day… And then Thursday and Friday, they are at the dad's house… that's the day I get to go out. I get to socialize. I get to sleep. I get to do this more room for, for me to, yeah.}{P1}.

Through her caregiving experiences, she developed her resilience leveraging intrapersonal assets–maintaining an awareness of what drains her or recharges her–to manage her emotional bandwidth for caregiving, \inlinequote{I discovered that the coaching recharges me as well....that keeps me going. I think just, yeah, being aware of like my sensitivities, like, okay, you cannot deal with too many difficult things. Then I'll be drained. Then I don't have enough space for the kids.}{P1}.

She also leverages social media tools for reflecting on her everyday experience, which she found especially useful as a substitute to reflecting with a spouse, \inlinequote{after divorce... you don't have anybody to really talk to about your kids…about the small things…So Instagram became my sort of that dumping, you know, the outlet for me to go like, hey, today this happened, this happened. And I don't care if people read it or not, it's more for myself.}{P1}

\medskip

\textbf{Case study 4}: P2 is a father of two–an eleven-months old infant son and a nine year old daughter. He works in an international MNC, where workplace policies and frequent travel shapes the company’s supportive work culture for flexibility where "everyone is okay not seeing you in the office”. As a manager level leader in his company, he has considerable agency over his time, \inlinequote{Till the moment I'm delivering the results doesn't matter how my day looks like because there are days when I'm in the call still even 2am or 3am right.}{P2}. 

At home, his main care responsibility is for his infant son, which he manages with his wife and a live-in domestic worker. As he travels frequently for work and is away for two to three weeks at a time, he plans closely with his wife, who also works in an MNC with supportive flexible work policy and culture, to manage their children’s care. They review work schedules together weekly, taking into account each other’s workload and his travel schedules to determine who will work from home to care for their child. As he shares, \inlinequote{between… three of us, we keep rotating our roles, right? So we'll decide that every week, whose week is heavy versus lighter. And accordingly, we will decide which days we'll go to the office versus home. So that we can also support the child.}{P2}.

They also leverage digital tools such as WhatsApp for their coordination and planning, creating dedicated chat groups to keep track of important dates, information and tasks, \inlinequote{...me and my wife, we also have created… WhatsApp groups in which just share with a different type… we have a WhatsApp group for the weekend activities… we always put you know the dates. So for example… my daughter's orchestra is coming, so that we don't miss out those dates… So we always keep check on that one. All the activities that we have to do over the weekend, for example, buy this stuff, ordering this machine, whatever the household chores, we keep a note in that chat group, we have a list so that we don't forget.}{P2}.

When it is his turn to work from home, he manages the day together with their live-in domestic worker, updating her of his schedule at the start of the day, \inlinequote{this was basically me and my domestic worker only who had to manage the entire day. So I generally also tell her that how my day looks like every day in the morning.}{P2}. Throughout the day, he tags-team with his domestic worker to juggle both work and care, handing care over to her when he has to attend meetings online, and taking over care when she runs errands, \inlinequote{the way we have divided the responsibility is that when he is sleeping, I will be sitting… working while keeping a constant eye on him. But if he wakes up, I'll again call my domestic worker to help him to go to bed again… But yeah, there are days when you know, my domestic worker has to go to the grocery store or you know, she's busy with some household chores, then yeah, we have to switch for for a partial activity.}{P2}. In the evenings, his wife takes over care after she returns from work, allowing him another block of focused work, \inlinequote{then there is a time for my wife also to come back home. And then she takes the lead and then I again focus on my laptop. That is typically the way my day goes.}{P2}.

\end{document}